\documentclass[preprint,12pt]{elsarticle}

\usepackage{amssymb}
\usepackage{xcolor}
\usepackage{soul}
\usepackage{tabularx} 
\usepackage{url}

\usepackage{breakurl}
\usepackage[breaklinks]{hyperref}

\journal{Information Sciences}

\begin{document}

\begin{frontmatter}


\title{A survey of the European Open Science Cloud services for expanding the capacity and capabilities of multidisciplinary scientific applications}


\author[inst1]{Amanda Calatrava\corref{correspondingauthor}} \cortext[correspondingauthor]{Corresponding author}
\ead{amcaar@i3m.upv.es}

\author[inst21,inst22]{Hernán Asorey}
\author[inst6]{Jan Astalos}
\author[inst3]{Alberto Azevedo}
\author[inst9]{Francesco Benincasa}
\author[inst1]{Ignacio Blanquer}
\author[inst6]{Martin Bobak}
\author[inst2]{Francisco Brasileiro}
\author[inst9]{Laia Codó}
\author[inst7]{Laura del Cano}
\author[inst5]{Borja Esteban}
\author[inst9]{Meritxell Ferret}
\author[inst8]{Josef Handl}
\author[inst4]{Tobias Kerzenmacher}
\author[inst5]{Valentin Kozlov}
\author[inst8]{Aleš Křenek}
\author[inst3]{Ricardo Martins}
\author[inst10]{Manuel Pavesio}
\author[inst20]{Antonio Juan Rubio-Montero}
\author[inst10]{Juan Sánchez-Ferrero}

\affiliation[inst1]{organization={Instituto de Instrumentac\'on para Imagen Molecular (I3M), Universitat Polit\`ecnica de Val\`encia},
            city={Valencia},
            postcode={46022},
            country={Spain}}

\affiliation[inst2]{organization={Federal University of Campina Grande (UFCG)},
            city={Campina Grande},
            country={Brazil}}

\affiliation[inst3]{organization={Laboratório Nacional de Engenharia Civil (LNEC)},
            city={Lisbon},
            country={Portugal}}

\affiliation[inst4]{organization={Institute for Meteorology and Climate Research--Atmospheric Trace Gases and Remote Sensing (IMK-ASF), Karlsruhe Institute of Technology (KIT)},
            city={Karlsruhe},
            country={Germany}}
\affiliation[inst5]{organization={Steinbuch Centre for Computing (SCC), Karlsruhe Institute of Technology (KIT)},
            city={Karlsruhe},
            country={Germany}}

\affiliation[inst6]{organization={Institute of Informatics, Slovak Academy of Sciences},
            city={Bratislava},
            country={Slovakia}}

\affiliation[inst7]{organization={Centro Nacional de Biotecnologia, CSIC},
            city={Madrid},
            country={Spain}}
\affiliation[inst8]{organization={MU},
            city={Brno},
            country={Czech Republic}}

\affiliation[inst9]{organization={Barcelona Supercomputing Center (BSC)},
            city={Barcelona},
            country={Spain}}

\affiliation[inst20]{
    organization={Centro de Investigaciones Energéticas, Medioambientales y Tecnológicas (CIEMAT)},
    addressline={Av. Complutense 40}, 
    city={Madrid},
    postcode={28040}, 
    state={Madrid},
    country={Spain}
}
\affiliation[inst21]{
    organization={Medical Physics Department, Comisión Nacional de Energía Atómica (CNEA)},
    addressline={Centro Atómico Bariloche}, 
    city={San Carlos de Bariloche},
    postcode={8400}, 
    state={Río Negro},
    country={Argentina}
}

\affiliation[inst22]{
    organization={Instituto de Tecnología en Detección y Astropartículas (ITeDA, CNEA/CONICET/UNSAM)},
    addressline={Centro Atómico Constituyentes}, 
    city={Villa Maipú},
    postcode={1450}, 
    state={Buenos Aires},
    country={Argentina}
}

\affiliation[inst10]{
    organization={Control, Observation and tracking systems, Space Management Area (Indra Sistemas SA)},
    addressline={Ctra. de Loeches, 9}, 
    city={Torrejón de Ardoz},
    postcode={28850}, 
    state={Madrid},
    country={Spain}
}

\begin{abstract}
Open Science is a paradigm in which scientific data, procedures, tools and results are shared transparently and reused by society as a whole. The initiative known as the European Open Science Cloud (EOSC) is an effort in Europe to provide an open, trusted, virtual and federated computing environment to execute scientific applications, and to store, share and re-use research data across borders and scientific disciplines. 
Additionally, scientific services are becoming increasingly data intensive, not only in terms of computationally intensive tasks but also in terms of storage resources. Computing paradigms such as High Performance Computing (HPC) and Cloud Computing are applied to e-science applications to meet these demands. However, adapting applications and services to these paradigms is not a trivial task, commonly requiring a deep knowledge of the underlying technologies, which often constitutes a barrier for its uptake by scientists in general. 
In this context, EOSC-SYNERGY, a collaborative project involving more than 20 institutions from eight European countries pooling their knowledge and experience to enhance EOSC's capabilities and capacities, aims to bring EOSC closer to the scientific communities. This article provides a summary analysis of the adaptations made in the ten thematic services of EOSC-SYNERGY to embrace this paradigm. These services are grouped into four categories: Earth Observation, Environment, Biomedicine, and Astrophysics. The analysis will lead to the identification of commonalities, best practices and common requirements, regardless of the thematic area of the service. Experience gained from the thematic services could be transferred to new services for the adoption of the EOSC ecosystem framework.
\end{abstract}

\begin{graphicalabstract}
\includegraphics[width = 1\linewidth]{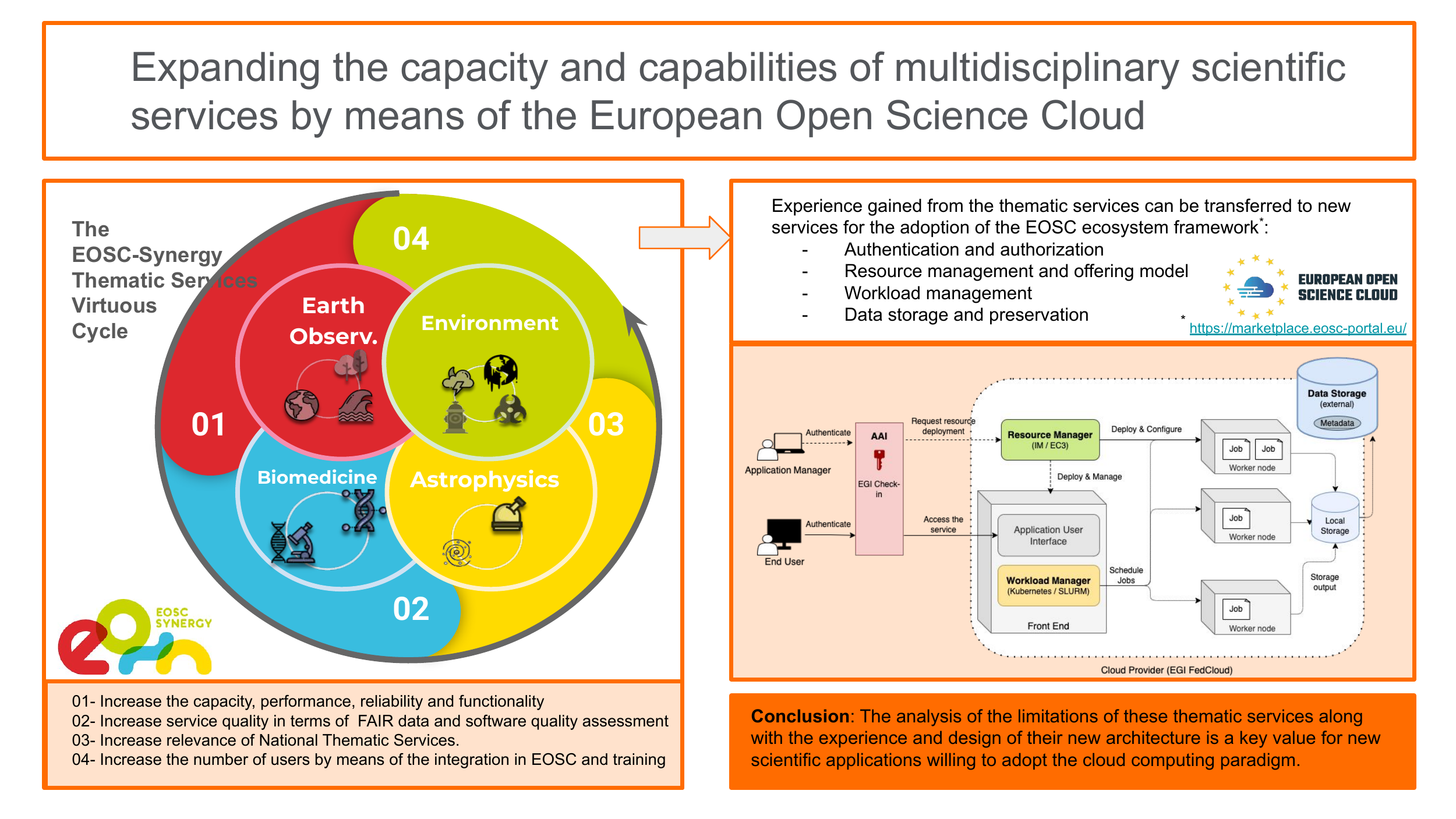}
\end{graphicalabstract}

\begin{highlights}
\item The European Open Science Cloud (EOSC) is an initiative aiming to offer a virtual environment for open access to services to store, share, process and reuse research data and other research digital objects, such as software.
\item The adaptation, improvement and quality assessment of thematic services on a Federated Data Infrastructure strongly aligns with the objectives of EOSC.
\item A key factor for the success of EOSC is performance and acknowledgement by the users.
\item We present an analysis of the adoption of services from the EOSC catalogue that provide feedback on the usability and relevance of the model.
\item The ten thematic services analysed provide different experiences on enhancing community-oriented data services to be released in an Open Science environment.
\end{highlights}

\begin{keyword}
Open Science \sep Cloud Computing \sep federated infrastructure \sep multidisciplinary \sep EOSC
\end{keyword}

\end{frontmatter}

\section{Introduction}
\label{sec:intro}

e-Science studies, enacts, and improves the ongoing process of innovation in computationally-intensive or data-intensive research methods \cite{ieee20212021}; typically this is carried out collaboratively, often using distributed infrastructures. Open Science \cite{opensciencedef} is the practice of science in such a way that others can collaborate and contribute, with research data, lab notes and other research processes freely available, under terms that enable reuse, redistribution and reproduction of the research and its underlying data and methods.

Scientific applications place higher demands on computing power every year. The need for large-scale computing resources, including specific hardware needs such as GPUs, together with the increasing demand for storage resources due to the large amount of data generated by these types of applications, are a challenge for both researchers and computer scientists. e-Science makes use of e-Infrastructures \cite{einfrastructures} which are collaborative virtual environments that provide digital services and tools to meet this resource demand.
 
 e-Infrastructures are based on distributed backends ranging from High-Performance Computing to Cloud Computing. However, adapting already existing software applications to these paradigms is not trivial \cite{blanquer2012requirements}. This process commonly requires an in-depth knowledge of the underlying technologies to truly take advantage of their benefits, as it usually requires refactoring the architecture of the application. This adaption becomes even more challenging on the Cloud computing paradigm, due to the complexity of virtualization and elasticity behaviour together with the vast range of services available and the variety of resource types. This fact can be a barrier for scientists and researchers from fields outside of computer science to adopt these solutions, thus complicating the path of research innovation.
 
 
 Another obstacle in adopting the e-Science and Open Science paradigms is the fulfilment of the FAIR \cite{fair_data_maturity_model_working_group_2020_3909563} (Findability, Accessibility, Interoperability and Reusability) principles, which imply permanent and discoverable identifiers for fully annotated data and metadata. Specifically, in Europe, we can find the European Open Science Cloud (EOSC) \cite{eosc}, a European initiative co-funded by the European Commission that aims to facilitate the deployment and consolidation of an open, trusted, virtual, federated environment in Europe to store, share and re-use research data across borders and scientific disciplines promoting open science practices and providing access to a rich array of related services.

As part of this initiative, the EOSC-Synergy \cite{eoscsynergy} project aims to increase the uptake of EOSC through the capacity and capability building using the experience, efforts and resources of national publicly funded digital infrastructures. The project has identified ten thematic services in four scientific domains (Earth Observation, Environment, Bio-medicine, and Astrophysics) to increase the uptake of EOSC services. These thematic services are heterogeneous and cover a wide range of requirements, maturity levels, user targets and usage models. 

The ten thematic services also provide helpful best practices for future new services to be developed or adapted to this environment, as they address challenges on federated Authentication and Authorization Infrastructures (AAI), elastic data processing, interoperability with data infrastructures, metadata management and accounting, that apply to many other applications in same or different scientific domains. The analysis of the limitations of these thematic services along with the design of the new architecture of the thematic services will be of particular interest for new scientific applications that want to embrace this paradigm.

In this work, we present each of the thematic services of EOSC-Synergy and then analyse the gaps and bottlenecks in terms of authentication and authorisation services, resource provisioning, workload management, and data storage. This section is followed by the analysis of the different tools and services provided in the EOSC Marketplace \cite{eosc_catalog} that can meet the needs of the thematic services. We then present the adoption of services, tools and technologies used by the thematic services to address their needs, and then model a generic scientific application that can be the starting point for new scientific thematic services. Next, to better illustrate the work done in the adoption of tools and services by the thematic services, we discuss the adoption issues in the service instantiation section. Finally, we analyze the state of the art to identify the most important work related to the implementation of thematic services in the e-Science paradigm, and conclude the paper with the main observations.

\section{Thematic Services in EOSC Synergy}
\label{sec:ts}
In this section we briefly describe each one of the ten thematic services of EOSC Synergy, grouped by scientific discipline. Moreover, the expected outcome of the integration with EOSC Synergy for each one of the thematic services are pointed out in this section.

\subsection{Thematic Services in Earth Observation}
In the field of Earth Observation, three thematic services deal with analysing large satellite imagery, from monitoring coastal changes and inundations, to estimating forest masses and crops. They are addressing different types of targets. Specifically, the three services are:

\begin{itemize}
    \item WORSICA (Water Monitoring Sentinel Cloud Platform) \cite{worsica_service1,worsica_service2}: A service for the detection of water using satellites, Unmanned Aerial Vehicles \& in-situ data. WORSICA can be used for coastline detection, inland water bodies detection and water leak detection on irrigation networks. WORSICA aims at integrating multiple-source remote sensing and in-situ data to determine the presence of water in coastal and inland areas. WORSICA enables the research communities to generate maps of water presence and water delimitation lines in coastal and inland regions. These products can be useful for emergency and planning methodologies in case of inundations or reservoir leaks. In the frame of EOSC, the service will be scaled up to a European level to reach all interested research communities. 
    \item SAPS (Surface Energy Balance Automated Processing Service) \cite{CUNHA2020104341}: Used to estimate Evapotranspiration and other environmental data that can be applied, for example, on water management and the analysis of the evolution of forest masses and crops. SAPS allows the integration of Energy Balance algorithms to compute the estimations that are of special interest for researchers in Agriculture Engineering and Environment. These algorithms can be used to increase the knowledge on the impact of human and environmental actions on vegetation, leading to better forest management and analysis of risks. SAPS is being developed in Brazil, but with the adoption of EOSC services it is expected to facilitate European scientists to exploit the evapotranspiration estimation services from remote sensing imagery.

    \item G-Core (Acquisition, cataloguing and processing EOS data) \cite{gcore_poster}, \cite{gcore_esa}, \cite{gcore_service}: G-Core is a production-ready technology used as a service at ESA’s and national programs that provides a Data Manager for spatial and non-spatial purposes and a framework for third-party processors. G-Core is a service for the acquisition, storing, cataloguing and processing data from several Earth Observing System (EOS) missions. Its two main functionalities are: i) a Data Manager for spatial and non-spatial purposes; and ii) a Processing framework to host external processors developed by third parties to generate added value products based on Satellite imagery. The main goal of its integration in the EOSC ecosystem is to offer the service as a Payload Data Ground Segment (PDGS) in the cloud for future ground segment space missions or as a processing framework to plug in different processors that can make use of the Copernicus resources or private data in order to produce different levels of products to be delivered to the users. Thus, the expected impact of the adaptation of the service is to democratize the usage of Earth Observation (EO) data out of the scope of nominal fields. It will help to define new products and services mixing Earth Observation data with other types of data for scientific and social environments.
\end{itemize}

\subsection{Thematic Services in Biomedicine}
In this area, the thematic services cover the benchmarking of Genomic data processing tools and the processing of Cryo-electron microscopy imaging. The services are:

\begin{itemize}
    \item SCIPION (CryoEM data processing for Structural Biology \cite{DELAROSATREVIN201693}): Cryo-Electron Microscopy Service is an image processing framework used to obtain 3D maps of macromolecular complexes using cryo Electron Microscopy. It has been developed as a plugin-based workflow management system that integrates many important software packages available in the field. The integration of Scipion with Cloud services allows users from the Instruct Research Infrastructure to deploy a dynamic cluster in the cloud to keep processing the data acquired at an Electron Microscopy facility. This cluster has all cryoem packages and software needed to obtain a 3D structure and is powered by EOSC compute resources on the back-end. This means that scientists with minimal computational background (or compute resources of their own) can access the latest tools as well as powerful computational resources to obtain a refined 3D structure to be published and shared with the community.
    \item OpenEBench \cite{OpenEBench} (ELIXIR \cite{elixir} benchmarking and technical monitoring platform): Used to evaluate Life Sciences research software, OpenEBench is an observatory for software quality based on the automated monitoring of FAIR for research software metrics and indicators. The OpenEBench platform supports both the technical monitoring of scientific software and scientific benchmarking activities carried out by Life Sciences Communities. Its architecture has three different engagement levels that allow communities at different maturity stages to make use of the platform. It also connects with ELIXIR Core Data Resources and Deposition databases to use data needed by the Scientific Communities activities. The expected impact of the integration with EOSC services is that Life Science researchers will have semantically annotated, up-to-date collections of benchmarked analytical workflows and tools, organized by scientific communities for specific topics, which can be deployed across heterogeneous systems.
\end{itemize}

\subsection{Thematic Services in Astrophysics}
In Astrophysics, the LAGO thematic service sets up a European service for the Latin American Giant Observatory. 

\begin{itemize}
    \item LAGO, the Latin American Giant Observatory~\cite{lagoproject, lagoprojectweb}, is an extended cosmic ray observatory, consisting of a wide network of water Cherenkov detectors currently deployed at 10 countries in Latin America, from the south of Mexico to the Antarctic Peninsula. The geographic distribution of LAGO allows the realization of diverse astrophysics studies at a regional scale. LAGO is mainly oriented to perform basic research focusing on three main areas: high energy phenomena, the measurement of atmospheric radiation at ground level and space weather and climate monitoring.
    All the LAGO analyses are supported by data-intensive computational frameworks, that integrate different simulations tools with own designed data-analysis codes to determine in a very precise way, the signals measured or expected at any detector of any type, in any particular site around the World, and under realistic atmospheric and geomagnetic time-evolving conditions. The final purpose of the LAGO Thematic Service~\cite{LAGO_WSC21} is to enable the universal profit and contribution of this research, within and outside LAGO Collaboration, through a sustainable Virtual Observatory and standardised computational model.
    
\end{itemize}

\subsection{Thematic Services in the Environmental domain}
Finally, in the area of Environment, the fourth group of thematic services include sand and dust storm forecasting, untargeted mass-spectrometry analysis for toxins,  water network distribution simulation and the monitoring of stratospheric ozone in climate models. Specifically, these services are as follows:

\begin{itemize}
    \item SDS-WAS (A Service related to the mineral dust forecast) \cite{sdswas, bdfc}: SDS-WAS is a service that aims at improving capabilities for more reliable sand and dust storm (SDS) forecasts. It supports institutional entities to warn about possible dust events and to foster the study of dust-related phenomena. The framework collects numerical model outputs and observational data from a wide set of worldwide partners plus internally developed. A wide set of post-processed analysis and statistics are generated, and results in form of plots, tables or numerical (binary) data are disseminated to a variety of users (e.g. public institutions, researchers). The integration of the framework in EOSC will increase the volume of data hosted and processed, to reach a wider set of end-users; improve compliance to data FAIR principles and reinforce the robustness of the whole service infrastructure.
    \item UMSA (Untargeted Mass- Spectrometry Analysis) \cite{umsa_service}: UMSA aims at processing mass spectrometry data to correlating the whole spectra (ie. all the present compounds) with other data (social, medical, other sample analyses, etc.) to work with more complex hypotheses on the impact of environment in human health. The data are unrecoverable, therefore long-term data storage is required, together with appropriate data curation. By means of the integration in EOSC, uniform access to data and computing resources are provided, scaling the service to the target European-wide user community. 
    \item MSWSS \cite{mswss_service} (Water Supply Systems modeling and analysis): MSWSS is a service for modeling and analyzing Water Supply Systems which integrates the analysis of toxins in drinking-water supply networks with water distribution network simulation. It allows water infrastructure operators and researchers to analyse hazardous events (e.g. toxin propagation within a pipe system) and may be used for preparation of risk management plans for water utilities. The integration with EOSC computing infrastructure and data sharing services will enable modelling more complex water supply systems and increase the number of scenarios for the analysis.
    \item O3AS (Ozone Assessment Service) \cite{o3as_service}: The O3AS service provides an invaluable tool to extract ozone (O$_3$) trends from extensive climate prediction model data to produce figures of stratospheric ozone and figures of dates by when depleted O$_3$ recovers to pre-ozone hole levels. This service is conceived to assist scientists in visualising ozone data from large climate models by calculating dates for the recovery of the ozone layer and providing trends of the ozone abundance in the atmosphere to produce results in the form of figures in publication quality. The integration of the service in EOSC has increased its capacity to process large volumes of data (in terms of TBs) and to facilitate the management of the complex workflow to generate key metrics. Climate model data has many fields of physical quantities. The relevant quantities for the ozone column calculation has to be extracted and processed in order to be visualized efficiently. A pre-processor is running on an HPC system reducing the large data set so that it can be read by an REST API to produce figures without noticeable delays. 
\end{itemize}

\section{Gaps and Bottlenecks analysis}
\label{sec:gaps}

Before starting the integration of the thematic services in EOSC, we have performed a deep analysis with each one of them to properly identify in advance the needs and requirements to increase the capacity, performance, reliability and/or functionality of the services. Thus, every thematic service has made an analysis of the technical services used for managing the users, computing and data. Table \ref{tab:gaps} shows a summary of the limitations, lacks and needs identified by the Thematic services of the project. 

{\setlength\extrarowheight{2pt}
\begin{table}[]
\resizebox{\textwidth}{!}{\begin{tabular}{|l|l|}
\hline
\begin{tabular}[c]{@{}l@{}}Thematic \\ Service\end{tabular} &
  Limitations and needs \\ \hline
WORSICA &
  \begin{tabular}[c]{@{}l@{}}- Improve download speed and number of concurrent downloads of satellite images.\\ - Increase storage of the images needed for the algorithm.\\ - Increase computational resources: GPU and RAM to speedup the image processing.\\ - Seamless authentication and authorization for end users.\end{tabular} \\ \hline
SAPS &
  \begin{tabular}[c]{@{}l@{}}- Need for a larger-scale deployment: computing, storage and data access.\\ - Scalability and standardisation of services\\ - Integrated and widely supported AAI\end{tabular} \\ \hline
GCore &
  \begin{tabular}[c]{@{}l@{}}- Overcome limited access to data repository due to network bandwidth restrictions.\\ - Infrastructure resources for processing and reprocessing large data sets.\\ - Data delivery volume. Increasing size of files to be delivered to users.\end{tabular} \\ \hline
SCIPION &
  \begin{tabular}[c]{@{}l@{}}- Insufficient Cloud resources for the workflow: GPUs, CPUs and RAM\\ - Need of a Resource Management able to optimize the use of cloud resources.\\ - Storage limitations and data transfer performance: 1-3 TB raw data.\\ - Distributed and shared file system.\end{tabular} \\ \hline
OpenEBench &
  \begin{tabular}[c]{@{}l@{}}- Need to work on heterogeneous systems to reach Life Sciences Communities\\ - Need to efficiently store processed data and workflows in a FAIR manner.\end{tabular} \\ \hline
LAGO &
  \begin{tabular}[c]{@{}l@{}}- Limitations on data preprocessing.\\ - Needs data storage that copes with FAIR, curation and harvesting; \\ - Need for computing power for simulations, together with optimal scheduling.\end{tabular} \\ \hline
SDS-WAS &
  \begin{tabular}[c]{@{}l@{}}- Lack of services needed for Data storage and curation. \\ - Lack of computing power for data analysis on-demand.\\ - Lack of reliability of data sources, especially about observations\end{tabular} \\ \hline
UMSA &
  \begin{tabular}[c]{@{}l@{}}- Long-term data storage is required, together with appropriate data curation.\\ - Tracking provenance of the secondary (derived) datasets.\\ - Need for reimplementing UMSA algorithms to deal with sparse data.\end{tabular} \\ \hline
MSWSS &
  \begin{tabular}[c]{@{}l@{}}- Needs data protection measures because of the usage of confidential data.\\ - The data has to be stored in a private storage only. \\ - Implement security policies to protect VMs.\end{tabular} \\ \hline
O3AS &
  \begin{tabular}[c]{@{}l@{}}- Requires larger storage resources, specially improving data availability \\ - Fast handling of big data\end{tabular} \\ \hline
\end{tabular}}
\caption{Analysis of the limitations and needs of the Thematic Services.}
\label{tab:gaps}
\end{table}
}

In a preliminary analysis of the results, several technical commonalities and differences have been identified. All thematic services share the importance of using a robust AAI compatible with the ones used by the target institutions. With respect to resource management, all services have the interest of dynamically provisioning processing resources, most of the cases on demand. However, thematic services have different needs: from a dynamic dedicated cloud backend to an elastic cluster that shrinks or grows according to the workload, and even need to access external High-Performance Computing (HPC) and High Throughput Computing resources for massive Batch jobs execution. Regarding job Management, most thematic services use batch queues (like SLURM \cite{slurm} Batch queues or Galaxy \cite{goecks11t}), which could be extended to support containerised jobs. The usage of Kubernetes to orchestrate microservices and job queues of containers is also considered.

The most challenging part is the management of data. Thematic services have identified important issues on transferring and accessing large volumes of data and require smart caching, advanced data transferring and massive persistent data storage. There are two main approaches to cope with storage, from deploying their own datastore, e.g. DATAVERSE \cite{King07} instance; to the integration of external Data Infrastructures, like EGI DataHub \cite{VILJOEN2016148} and EUDAT \cite{lecarpentier2013eudat}. Moreover, services need to ensure the compliance to the FAIR principles to facilitate the access, cataloging and reusage of the data generated by their services. Thus, a storage service able to manage, together with the data,  metadata and unique data identifiers will be required.

Note that not all the services have identified gaps in all the previous aspects so each thematic service will focus the adaptation in the aspects that are more relevant according to the bottlenecks. Solutions available in the EOSC marketplace will be studied and prototyped in the next section before adapting them into the thematic services. 

\section{Analysis of the EOSC Portal Catalogue and Marketplace }
\label{sec:EOSCPortal}
In this section, our goal is to identify the key EOSC tools and services that can address the issues and needs analyzed above. Considering the gaps and bottlenecks identified, this analysis also considers potential alternatives of the current technical services used by the thematic services to overcome such issues. 

The EOSC Portal Catalog \& Marketplace \cite{eosc_catalog} has been developed from the perspective of users, identifying the needs to be supported and facilitating all the actors involved in implementing an open approach to science in a sustainable way. The catalog has more than 320 entries registered by the end of 2021, covering resources from several categories. According to a functional perspective, we can organise them into six categories:

\begin{itemize}
    \item Access physical \& eInfrastructures: offering generalist resources like virtual machines and containers as well as storage and network transport connectivity. By the end of 2021, 62 resources are listed under this topic. This category include Compute resource providers, workload managers, Resource orchestrators and data providers. Some of the services were thematic (e.g. discipline-specific). We identified 6 generic services that could address the requirements of the thematic services: B2SAFE for long-term data preservation, EGI Cloud Compute to provide IaaS cloud resources, EGI DataHubto provide online cloud storage resources, EGI High-Throughput Compute for batch workloads, EGI Workload Manager to orchestrate multi-site batch resources, and Infrastructure Manager - IM  to deploy virtual infrastructures on top of cloud offerings), according to the following criteria: Generic purpose, interoperability and support. A brief description of the services is provided next:
    \begin{itemize}
        \item EGI Cloud Compute \cite{EGICLOUD}, an IaaS from the EGI Federated Cloud that enables the user to deploy and scale virtual machines on-demand.
        \item EGI HTC Compute \cite{EGIHTC} enables running computational jobs at scale on the EGI infrastructure, which is provided by a distributed network of computing centres and offers more than 1,000,000 cores of installed capacity, supporting over 1.6 million computing jobs per day.
        \item EGI Workload Manager \cite{EGIWM}, a service to manage and distribute your computing tasks in an efficient way while maximising the usage of computational resources.
        \item EGI DataHub \cite{EGIDATAHUB}, a service that brings data close to the computing to exploit it efficiently and can be used to publish a dataset and make it available to a specific community or worldwide across federated sites.
        \item B2SAFE \cite{B2SAFE} is a service for long-term preservation of the EUDAT Data Collaborative Infrastructure, one of the largest e-infrastructures in Europe offering permanent storage capacity and integrated management services for research communities. EUDAT also provides other services such as B2SHARE, B2FIND, B2ACCESS. 
        \item Infrastructure Manager (IM) \cite{Caballer2015}
    \end{itemize}
    \item Aggregators \& Integrators , where we can find several tools and utilities to facilitate the access to services and resources, by means of indexing and annotation. Out of the 22 resources available in this category, we identified the Dynamic DNS service\cite{DYNDNS}, to easily add a DNS name to an instance deployed in the virtual infrastructure of EOSC, the EGI Fedcloud client \cite{EGIFedCloudCli}, that facilitates the access to the federated cloud computing platform, and B2FIND \cite{B2FIND}, another EUDAT's service, to annotate research objects, considering the requirements of the thematic services and similar criteria as in the first item.
    \item Processing \& Analysis mainly aimed at facilitating the management of computational resources and the scheduling the execution of workloads. Despite this is the category that accumulates the highest number of services, most of them are discipline-specific. Moreover, some of the resources where already listed under the first item. Here we can find Elastic Compute Clusters in the Cloud (EC3) \cite{CALATRAVA201613}, a tool to deploy virtual elastic clusters on top of IaaS clouds, and B2Handle (another EUDAT's service) to provide persistent identifiers to resources.
    \item Security \& Operations aims at guaranteeing that the overall system and the services operate securely and according to standard. In this case, thematic instances for authentication and authorisation may be preferred as researchers in the community already have acquire credentials. For this purpose we identify the EGI Check-In \cite{EGICHECKIN} service, the B2ACCESS \cite{B2ACCESS} and EduTeams \cite{EDUTEAMS}, along with ELIXIR AAI \cite{ELIXIRAAI} which is not listed in the catalogue.  
    \item Sharing \& Discovery relates both to services that produce data relevant to specific disciplines and horizontal services for data deposit and annotation. Only the service B2SHARE to enable sharing and publishing research data is considered. 
    The catalogue also includes an instance of Dataverse, although we have decided to deploy our own instance. The Dataverse Project, developed by the Harvard's Institute for Quantitative Social Science (IQSS), along with many collaborators and contributors worldwide, is an open-source web application to share, preserve, cite, explore, and analyze research data. 
    \item Training \& Support aims at facilitating the access to high quality technical information and tailored training materials. Services in this category are not considered. 
\end{itemize}

All these services have two different access modalities:
\begin{itemize}
    \item Direct access. This model is used by services which are instantiated up-front and which do not require intensive access to resources, or resources are provided directly by the user (e.g. EGI Check-in or Infrastructure Manager). Users are automatically forwarded to the service endpoints.  
    \item Access through orders. This model is used in services that require a non-trivial amount of resources (e.g. EGI Cloud Compute or B2SAFE). In this case, the user normally has to choose between different offerings, which may end up into costs.
\end{itemize}


\section{Adoption of EOSC services}
\label{sec:eosc_adoption}
As we have shown in section \ref{sec:gaps}, the ten thematic services have complementary requirements and features. However, in general they share needs on four different categories:

\textit{Authentication and Authorization Infrastructure (AAI)}. All services require users to be authenticated and authorised. In some cases, there is a need for delegation from the users that access the platform for accessing data or processing resources. In those cases, it is mandatory to have a coherent single-sign on mechanism. Other cases may require an AAI linked to popular scientific IdPs and implement the authentication via Virtual Organization membership. From the tools and services we identified from the TSs, EGI Check-in has revealed to be a widely accepted choice. Another option analyzed is B2Access, mainly for interacting with the infrastructure. There are also few cases in which users will use federated credentials to access the services - mainly related to storage. 

\textit{Workload Management}. Most of the cases deal with the execution of a set of batch jobs. In those cases, workload managers should be integrated to better take advantage of the computing resources. This will provide the capability to deal with a larger capacity and larger workloads. The options here range from using a standard batch queue (SLURM), that can be eventually powered up with automatic elasticity, to the usage of Kubernetes for the orchestration of a container-oriented approach. 

\textit{Resource Management}. Most of the thematic services require deploying a virtual infrastructure where the services that provide the functionality and the processing will take place. In most cases, the use of Infrastructure Manager (IM) or Elastic Compute Clusters in the Cloud (EC3) client have been identified by most of the thematic services as a technology capable of filling in this gap. Both tools could provide the capability of defining a virtual infrastructure as code and deploying it on the cloud. IM (for static infrastructures) or EC3 (for dynamic infrastructures) together with recipes for K8s, Slurm or Galaxy clusters on top of a dynamic dedicated cloud backend is the preferred solution. Moreover, some thematic services require links to external HPC resources (like Marenostrum in BSC) and HTC resources (EGI HTC compute) for the execution of massive Batch jobs. 

\textit{Data Storage}. The services need to have a storage connected to the processing that can be efficiently accessed. In this case, there is a wide range of different solutions proposed or implemented in the thematic services, ranging from external solutions like EGI-DataHub, B2SAFE and B2SHARE to local solutions based on Nextcloud, Dataverse, Elasticsearch and WebDav, where tipically the resource manager will be also in charge of deploying and configuring their own Datastore instance (e.g. DATAVERSE instance). 

{\setlength\extrarowheight{2pt}
\begin{table}[]
\begin{tabular}{|p{0.19\textwidth}|p{0.19\textwidth}|p{0.19\textwidth}|p{0.19\textwidth}|p{0.19\textwidth}|}
\hline
\textbf{Service} & \textbf{AAI} & \textbf{Workload Mng.} & \textbf{Resource Mng.} & \textbf{Data Storage} \\ \hline
\textbf{WORSICA} & EGI Check in & ArcCE, Batch (SLURM) & IM (TOSCA) & Nextcloud, Dataverse \\ \hline
\textbf{G-Core} &  CAS User/pwd \& EGI Check in  & GCore+ K8s & IM / EC3 & ElasticSearch \\ \hline
\textbf{SAPS} & EGI Check in & K8s & IM / EC3 & OpenStack Swift \\ \hline
\textbf{Scipion} & EGI Check in & Batch (SLURM) & IM / EC3 & Local + EGI DataHub \\ \hline
\textbf{OpenEBench} & Life Sciences AAI & WfExS + NextFlow & OpenNebula & Local + B2SHARE \\ \hline
\textbf{LAGO} & eduTEAMS + EGI Check-in & Batch (SLURM) & Local clusters +  IM / EC3 & EGI DataHub  ONEDATA \\ \hline
\textbf{SDS-WAS} & B2ACCESS & Batch (SLURM) & Local clusters & B2HANDLE / B2SAFE \\ \hline
\textbf{UMSA} & EGI Check in \&  Life- science AAI & Batch (SLURM) in  IM/EC3 (in Galaxy) & IM / EC3 & Local + S3 \\ \hline
\textbf{MSWSS} & EGI Check in &  Batch (SLURM) in EC3 (in Galaxy) & IM / EC3 & Local +  Dataverse \\ \hline
\textbf{O3AS} & EGI Check in & Batch (SLURM)  \& K8s & Local cluster + IM & Local + WebDAV \\ \hline

\end{tabular}
\caption{Adoption of technologies for each Thematic Service}
\label{tab:technologies}
\end{table}
}

Table \ref{tab:technologies} summarizes the selection of tools and services performed by each thematic service for the fourth categories detected. As a summary, three different (although compatible among them) AAI methods have been integrated (EGI Checkin, B2ACCESS, Life-Sciences AAI and eduTEAMS). Job scheduling ranges from solutions based on containers (using Kubernetes) to solutions using batch queues (mainly based on SLURM), supported in some cases by workflow frameworks such as Galaxy and instantiated through EC3. For the interaction with cloud resources, TOSCA \cite{Binz2014} and RADL recipes have been developed for Infrastructure Manager. Finally, data access is performed through different solutions such as Dataverse, EGI DataHub OneData, B2SHARE and B2SAFE, which clearly states the complexity of the data management issue and the wide range of solutions.

To sum up this section, with the adoption of all the services and technologies depicted in \ref{tab:technologies}, thematic services have experienced all these improvements:
\begin{itemize}
    \item Integration of standardized AAI IdPs to facilitate user management.
    \item Improvement of processing backends by replacing single computing instances with batch job queues, container management platforms or clients to high-throughput computing backends.
    \item Publishing the output results in persistent repositories.
    \item Improving repeatability and platform-agnosticism by describing the application topologies as code using standard TOSCA language.
    \item Self-management of resources to reduce maintenance costs. 
    \item Persistent Identifiers (PID) annotation of output data and integration in official harvesters.
\end{itemize}

Thanks to the rich analysis of the experience of these ten thematic services in the adoption of several tools, services and technologies to improve and solve their needs, the path to follow for a new scientific use case is far more easy. However, to clarify even more this process, and to easily identify the key services and technologies selected, we present in next section a generic application integrated in the EOSC ecosystem.

\section{Application Modelling}
\label{sec:appmodels}

This section uses as input the experience of the ten thematic services of EOSC-SYNERGY to define a canonical generic application architecture leveraging the services identified in the EOSC Marketplace catalogue in section \ref{sec:EOSCPortal}. Thematic services that have similar requirements as those described in section \ref{sec:gaps} can use as basis this architecture that relies on several tools and services from the EOSC ecosystem, together with well-known frameworks and technologies of the cloud computing paradigm, all of them carefully selected taking into account the selection made by the thematic services.

\begin{figure}[h]
	\begin{center}
	\includegraphics[width = 1\linewidth]{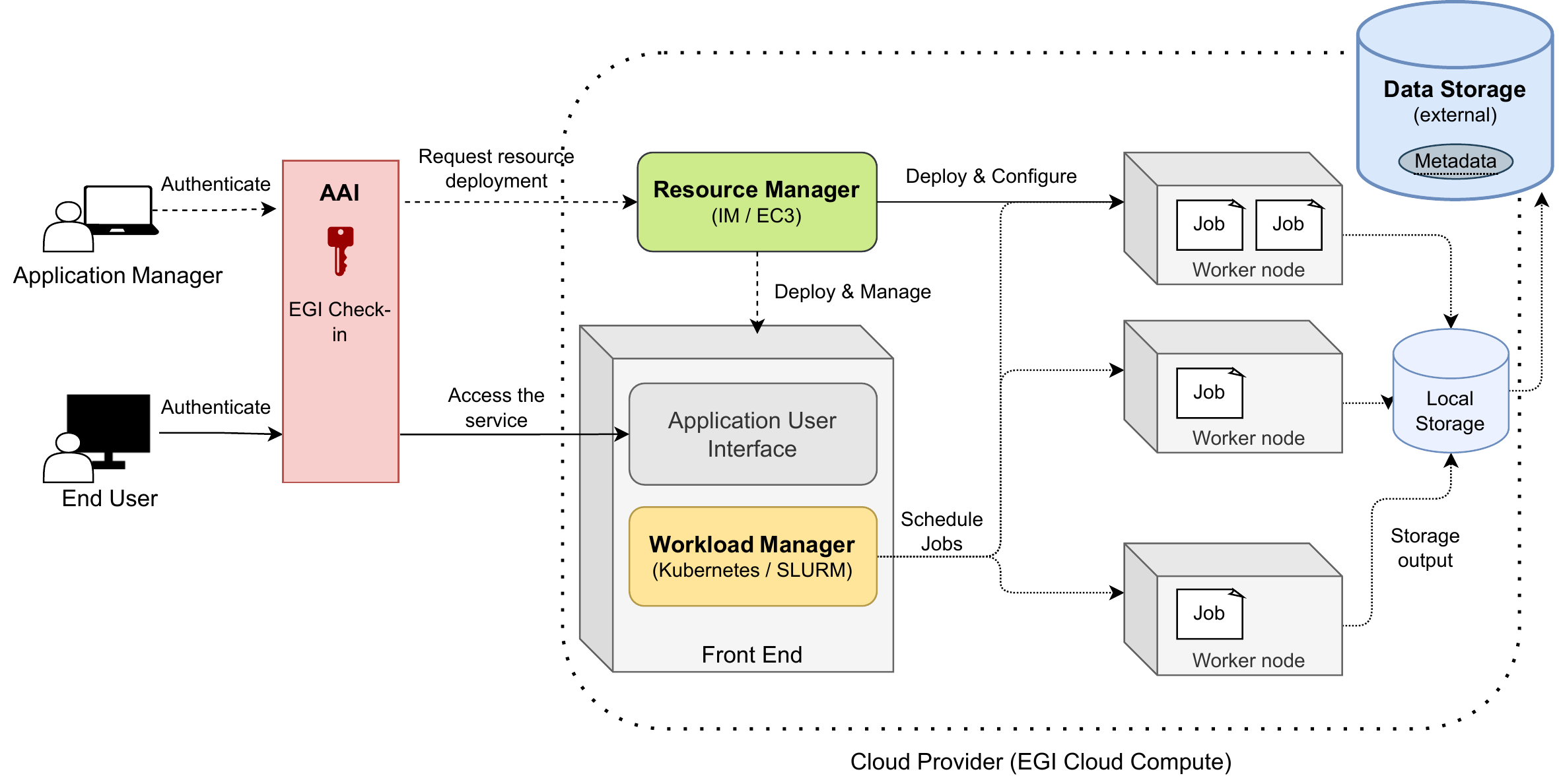}
	\end{center}
	\caption{\label{fig:arch_model}Architecture of the proposed solution. }
\end{figure}

First of all, after analyzing our ten use cases, we identified two different deployment scenarios: i) a single instance of the service shared by several users or communities, offered as a web portal able to manage users, data access, processing and visualization, supported by a shared or dedicated pool of resources (e.g. WORSICA or OpenEBench); and ii) an instance of the service deployed on demand, where each user deploys his/her own instance of the service on Cloud resources based on a combination of TOSCA recipes with Docker containers (e.g. SAPS, SCIPION).

Regardless of the approach chosen, Figure \ref{fig:arch_model} shows the architecture of this generic application that relies its deployment on top of the EGI Cloud Compute platform. The first layer that has been considered as essential by the thematic services is the authentication and authorization infrastructure. For that, EOSC offers the EGI Check-In service, that has been the most popular solution adopted by the thematic services. This service can be easily integrated with a service that is exposed to users through a web portal, and it will be used by both the application manager and the end-users to access the EOSC resources and the service itself. Once the user has been properly authenticated, and depending on the usage model that the service wants to use, he/she will access to the Application User Interface of the scientific application itself or to the portal of the resource manager that will facilitate the deployment of the scientific application instance. In the second scenario, the user will be redirected to the portal of IM or EC3 to deploy a virtual cluster configured on demand for his own usage. The selection between IM or EC3 has to be taken depending on the needs in terms of resource consumption. If elasticity is required, the tool to be used will be EC3. Otherwise, IM is the best tool to provide a static infrastructure configured on demand. Both solutions will require the preparation of a recipe where the application manager specifies the required steps and commands to properly install, configure and deploy the scientific service, together with the credentials to access the cloud provider.

In order to take advantage of the virtual infrastructure where the scientific application is running, we need to rely on a workload manager. From the analysis we have made, we have identified two different approaches: i) a traditional batch job queue, managed by the well-known SLURM scheduler; or a solution based on the containerization of jobs, where kubernetes has proven to be the most popular scheduler. Both options are feasible, depending on the approach that the scientific service wants to follow. However, the adoption of one of these workload managers might require an effort adapting the architecture of the tool to it, so this duty has to be consciously analyzed.

Finally, for the data storage, our recommendation is to use a solution that supports metadata to comply with FAIR principles, i.e. to make data Findable, Accessible, Interoperable, and Reusable (like Dataverse or B2SHARE). No matter if the storage solution will work locally or it will be an external service, one of the most important aspects is the support to metadata to properly index and facilitate reusing the data generated by the services, specially if this will be of interest to other researchers of the area.

\section{Service instantiation}
\label{sec:instanciation}
In this section we want to exemplify how the adoption of the new EOSC tools and services can address the gaps and bottlenecks detected by some of the project's thematic services. Specifically, we present seven examples from seven different scientific services showing the integration in each one of the categories described in section \ref{sec:eosc_adoption}. We have omitted the integration with the authentication and authorization service of EOSC (EGI Check-in), because this is a well-known process that is properly documented \cite{checkin}. Next subsections cover the cases for the rest of categories.

\subsection{Workload Management}
During the analysis of the thematic services, we have detected two main needs for the workload management: i) services that use a more traditional approach relying on local resource management systems based on queues of jobs; and ii) services that encapsulate the tasks in a container and rely on a container-based management system. To exemplify these two models, we analyze both Scipion and O3AS.

\subsubsection{Batch job oriented}\label{Scipion:Slurm}
The Scipion service aims to facilitates the life to  users from the Instruct-ERIC \cite{InstructERIC} Research community for processing their electron microscope data. Users who obtained her data through an Instruct granted project in an Electron Microscopy facility can request the use of the Scipion service by contacting the Instruct Image Processing Center (I2PC). Then, the request is reviewed and if the available quota permits it, the I2PC administrator will deploy a cluster using the Infrastructure Manager (IM). The user will then receive an email with instructions on how to access the front-end node and how to copy the data to start processing. She will be able to use the service for a maximum time of one month although if no other service requests are pending this time might be extended. The cluster will be destroyed by the I2PC team after the granted period finishes, giving the user enough time to download her results.
\begin{figure}[h]
	\begin{center}
	\includegraphics[width = .9\linewidth]{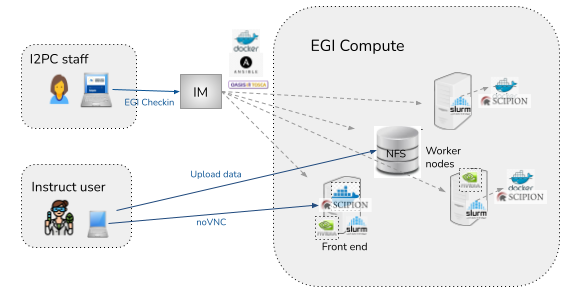}
	\caption{\label{fig:scipion-service}Scipion service usage and architecture.}
	\end{center}
\end{figure}


As shown in Fig.~\ref{fig:scipion-service} the cluster is defined in a TOSCA description that includes Ansible recipes and docker images to launch and configure the different nodes in the cluster. This TOSCA description is deployed through the IM. Once the cluster is running the user will be able to transfer the data to the front-end node and access the service using either a VNC client or NOVNC through a web browser.

The front-end host node runs a SLURM master and a Docker container that includes Scipion and related Cryoem packages configured to launch their jobs through SLURM. Once a job is sent to the queue a Docker container is run in one of the worker nodes to run the Scipion command. 

Hardware resources in which the cluster is deployed are part of the EOSC EGI Cloud Compute service that control access through Virtual Organizations (VO). In the case of the Scipion service deployment is only granted to members of the cryoem.instruct-eric.eu VO. 
The cluster shared storage is currently based on a local Ceph disk. 

\subsubsection{Container based}
O3AS (using Kubernetes) 

\begin{figure}[h]
	\begin{center}
	\includegraphics[width = .9\linewidth]{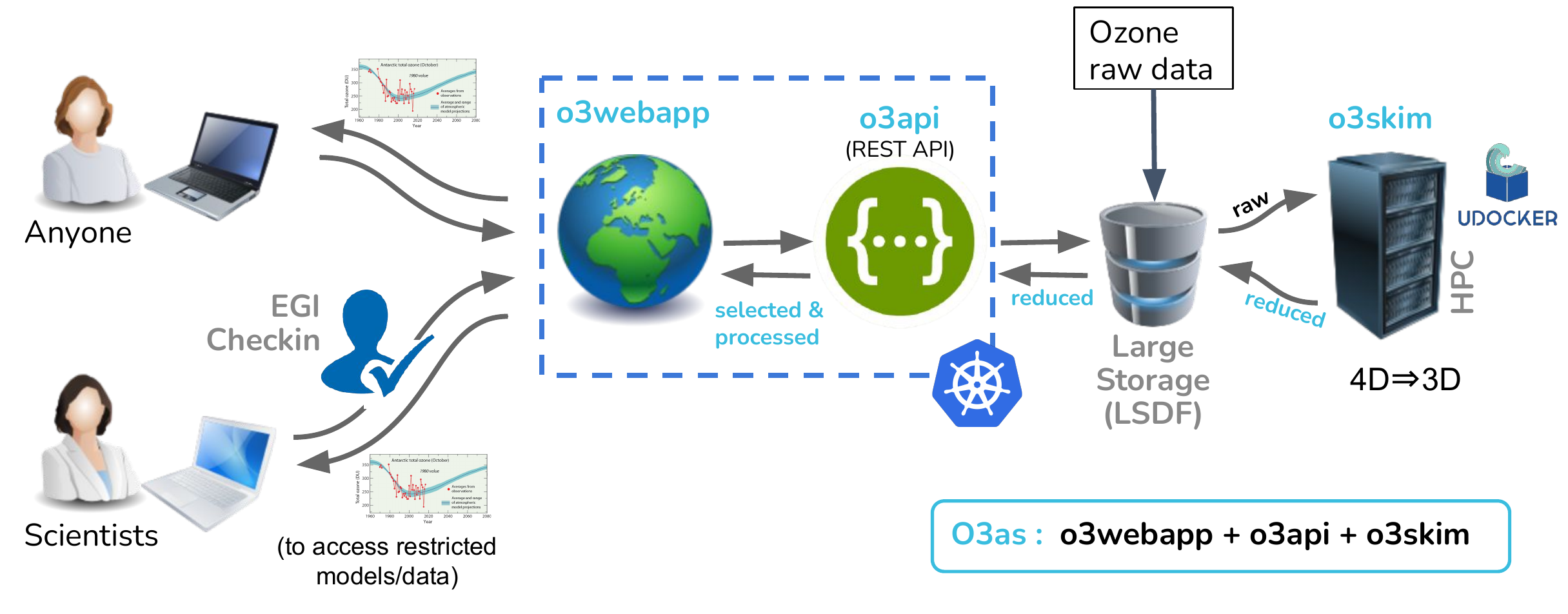}
	\caption{\label{fig:o3as_workflow}O3as service overview: it consists of three main components: o3webapp (\textit{to come}), o3api, and o3skim.}
	\end{center}
\end{figure}

O3as service is composed of three main components (Fig.~\ref{fig:o3as_workflow}): o3skim to reduce the original data to the parameters of interest, o3api to provide an API-based access to the skimmed data, and o3webapp for a user-friendly web interface (in development). O3skim runs on the local HPC system, while o3api and o3webapp are deployed in the Kubernetes system for the container scalability ensuring a fast enough response to user requests. If no Kubernetes system is available for the service providers, it can easily be instantiated by the means of Infrastructure Manager (IM). Then the following steps are applied:
\begin{enumerate}
    \item Install and configure cluster\_issuer to handle Certificates for secure HTTP connections, e.g from LetsEncrypt.
    \item Initialize Ingress resource to route external traffic.
    \item Deploy o3api component as a container with the pre-configured PriorityClass.
    \item Add Horizontal Pod Autoscaler (HPA) to respond with more containers on higher than usual loads.
    \item Finally, instantiate o3api as a service.
\end{enumerate}
If an already existing Kubernetes cluster is used instead, steps 1-5 have to be adjusted accordingly.

\begin{figure}[h]
	\includegraphics[width = .5\linewidth]{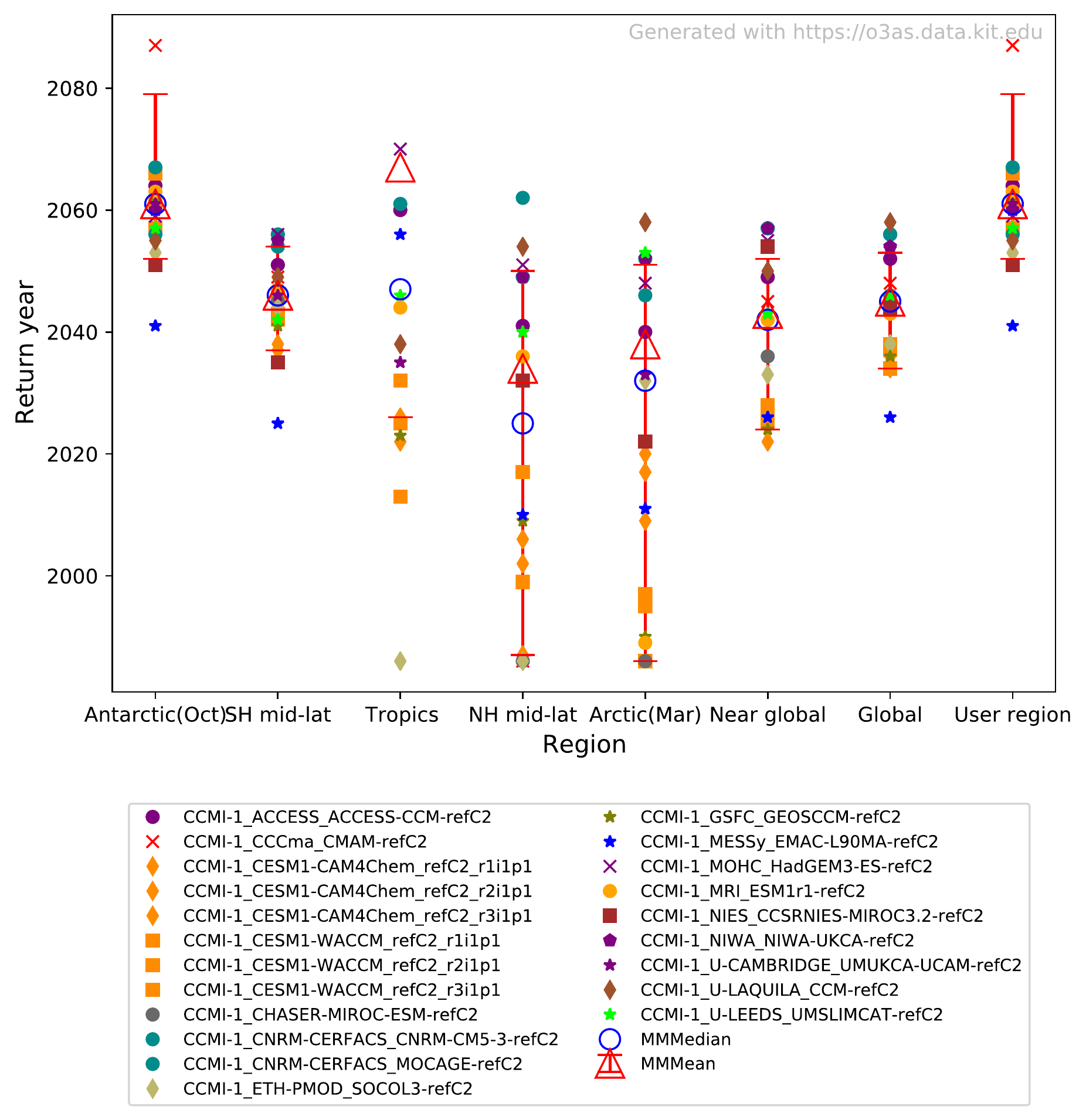}
	\includegraphics[width = .5\linewidth]{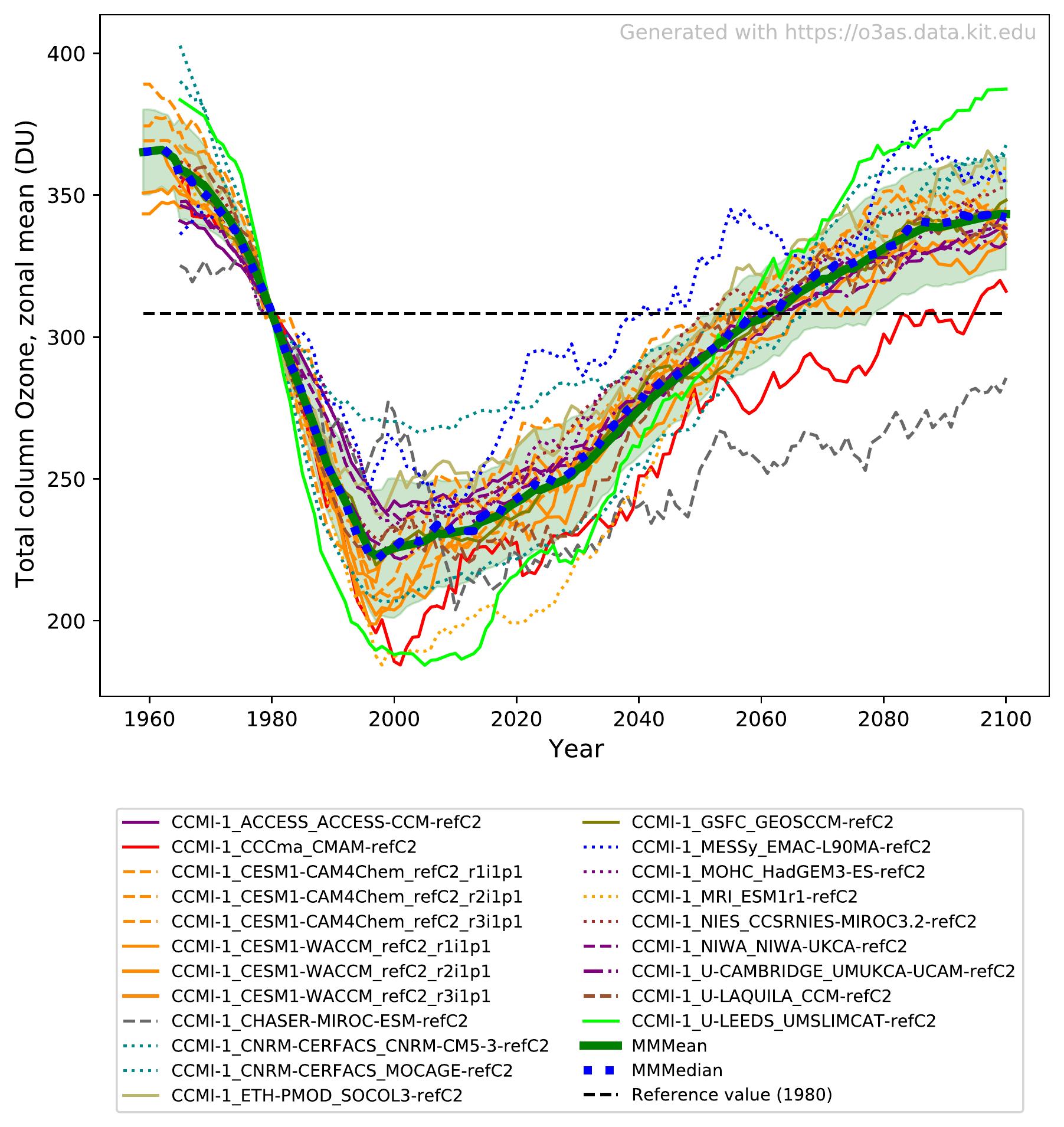}
	\caption{\label{fig:o3as_reults}Left: O$_3$ return dates for a recovery to ozone values in 1980. Right: Timeseries of total O$_3$ column data showing a decline of O$_3$ and the subsequent recovery of O$_3$ in the future.}
\end{figure}

The output of the services is a set of projections of the ozone distribution. Example figures for climate models with projections for ozone until the year 2100 are shown in Figure~\ref{fig:o3as_reults}.  

\subsection{Resource Management}
Regarding resource management in cloud computing providers, we have identified two different approaches: i) static infrastructures, where the number of nodes that compose the platform remains constant during the lifetime of the service; and ii) dynamic infrastructures, where elasticity models are applied to the platform to adapt its size to the workload, thus allowing to reduce costs and wasting resources. As in the previous section, we have chosen two thematic services to exemplify the adoption of these two different solutions.

\subsubsection{On-demand fixed infrastructures}\label{LAGO:fixed}

The flexibility for choosing the computing platform was one of the objectives of LAGO TS. Beyond the elasticity or the automatic management, the priority is providing resources that accomplish the needs of every specific calculation, and so these requirements may be as variable as its parametrization. Some of these simulations may face intensive requirements, such as scratching up to several TBs of data; accessing to many files through Internet;  continuously processing data in batch mode; or even sporadic calculations for demonstration purposes and scholars. Public clouds such as EOSC EGI Cloud Compute can tackle many of these tasks, but require an upfront reservation of the resources by demand and fix their environment. 
To be able to face all this different approaches, three different services were integrated: the Infrastructure Manager (IM) service, the software encapsulation in standardised Docker images and the profiting of the ubiquity of the OneData cloud storage (EGI DataHub sevice). These technologies allow dynamically instantiating the virtual infrastructures needed, which are maintained fixed over days or weeks~\cite{LAGO_WSC21}.

\begin{figure}[h]
	\begin{center}
	\includegraphics[width = 1.0\linewidth]{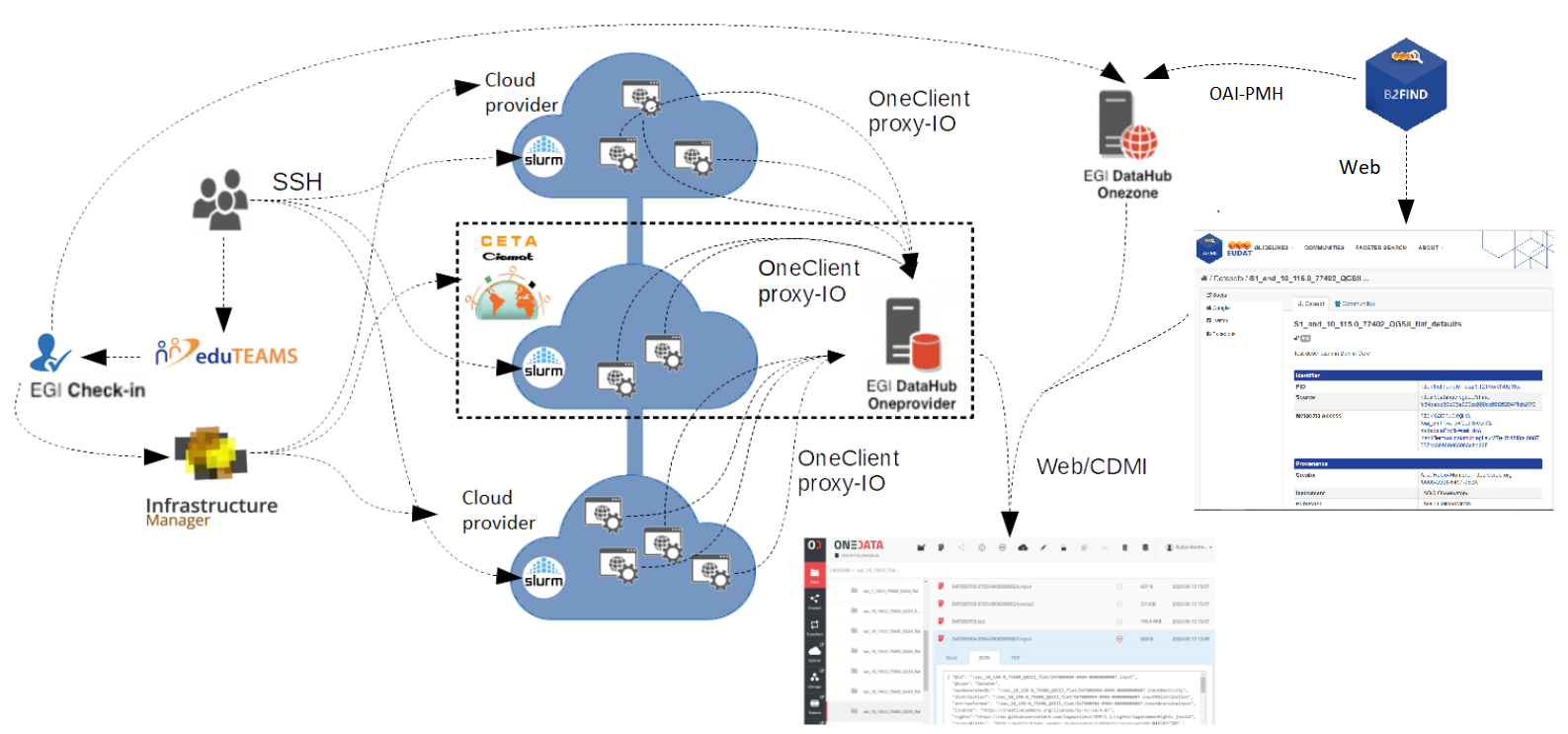}
	\end{center}
	\caption{\label{fig:LAGO-Chall} On-demand deployment of fixed   infrastructures and running LAGO software.}
\end{figure}

The resulting architecture is shown in Figure~\ref{fig:LAGO-Chall}. To fix a temporal infrastructure on public clouds, researchers dynamically apply for virtual machines or batch clusters through the IM service. Users can create any kind of cluster following, not only their own preferences, even the needs of specific calculations. In this sense, SLURM was a good choice as it is commonly used by many LAGO collaborators and the behaviour could be similar to those used by other TS, like, e.g., Scipion, as it is described in the subsection~\ref{Scipion:Slurm}. However, there are scientists would rather prefer to manage other implementations, such as Kubernetes, more suitable for Docker instances. Additionally, when tasks require scratching at the order of TBs, the user would not be allowed to spend the space accumulated by several computing nodes. In these cases, single virtual machines, are the most suitable choice. All these infrastructures are deployed by IM after a few clicks in its website. 

On the other hand, calculations can be arbitrarily performed by researchers by running the official LAGO docker images stored at DockerHub, which are periodically released by the CD/CI pipeline built on the JePL service \cite{jepl}. Thanks to the virtualised approach, the software encapsulated in these images can be run on any platform supporting Docker. However, as the FAIR paradigm has to be fulfilled, all the LAGO software is always bound to the OneData cloud storage (DataHub) and it comply with the AAI procedure for the LAGO VO in every run. Thus, independently the computing platform, results are always identified by PIDs and they are browsable at the DataHub portal and findable by public harvesters such as B2FIND. Therefore, the on-demand provisioning of adaptable infrastructures supporting Docker through the IM service, jointly the cloud storage via DataHub, allow users accomplishing their research without depending on other services.

As an example, we deployed a SLURM virtual cluster counting on 10 nodes with 16 Intel Xeon E7 cores and 250 GB of shared memory and disk. Then, we simulated the expected flux of the atmospheric radiation during the interaction of cosmic rays with the Earth atmosphere for every LAGO detector~\cite{Asorey2018}. We computed from $1$ to $7$\,days of the expected flux at high altitude or antarctic sites, reaching up to $1$ year of the energetic flux that is accounted for volcanic risk studies. These new integration times is an enormous statistical improvement when it is compared with previous results obtained in LAGO. Note that, e.g., a 24-hour flux in one of the high latitude (Antarctic) detector involves the simulation of $\sim 1.9\times 10^9$ different cascades. Thus, we simulated the impressive figure of $>10^{12}$ particles spending $>300$\,kCPU-hours, generating $>5$\,TB of synthetic data and metadata. 

\begin{figure}[!ht]
  \centering
  \includegraphics[width=0.45\textwidth]{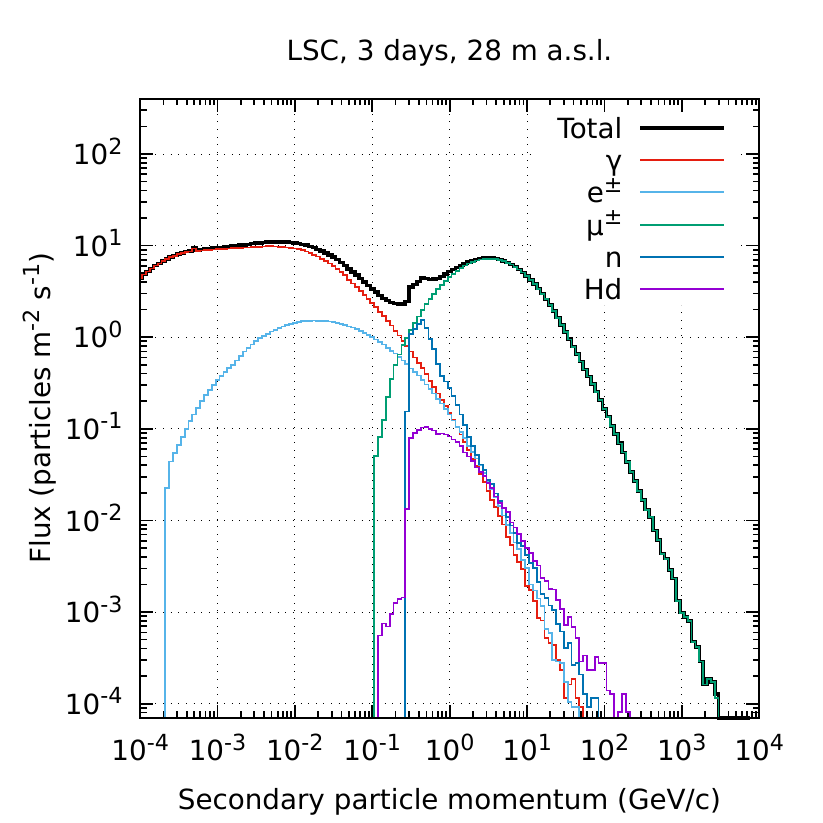}
  \includegraphics[width=0.45\textwidth]{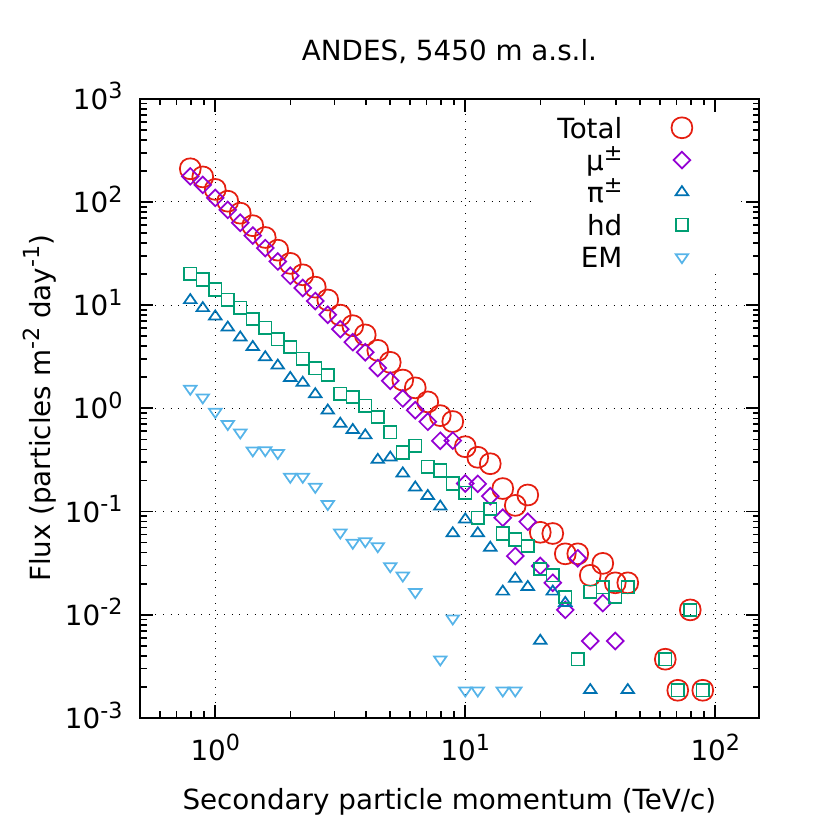}
  \caption{\label{fig:LAGO-results} Left: the energy spectrum of the flux of atmospheric radiation expected at La Serena (LSC) detector in Chile (at sea level) is used for designing, characterizing and calibrate new detectors and sites. Right: the expected flux of $>$TeV/c particles reaching the summit ($5450$\,m a.s.l.) of the mountain where the ANDES underground laboratory will be installed. Since these particles are capable to traverse up to thousands of meters of rock, being the background signals for neutrino physics experiments and dark matter searches.}
\end{figure}

Results shown in Figure~\ref{fig:LAGO-results} allow estimating radiation doses at different altitudes, which are currently used for designing new detectors; shielding instruments (e.g., HPC facilities); calculating the reference HEP flux for underground laboratories, volcanic risk assessments and mining prospecting~\cite{LAGO-icrc2021}.

\subsubsection{Elastic infrastructures}
 For cloud applications with varying load the static infrastructures might end up with resource waste. In order to adapt to the dynamic demand for computing capacity, the MSWSS service uses Elastic Cloud Computing Cluster (EC3) tool \cite{CALATRAVA201613} to create an elastic virtual cluster on top of the EOSC computing resources. EC3 CLI tool provides a set of pre-defined templates which can be combined and customised. It also allows to define custom templates with integrated Ansible scripts. This is used to create a template with strengthened security settings and additional configuration commands specific for the MSWSS service.
 
 \begin{figure}[h]
	\begin{center}
	\includegraphics[width = 0.65\linewidth]{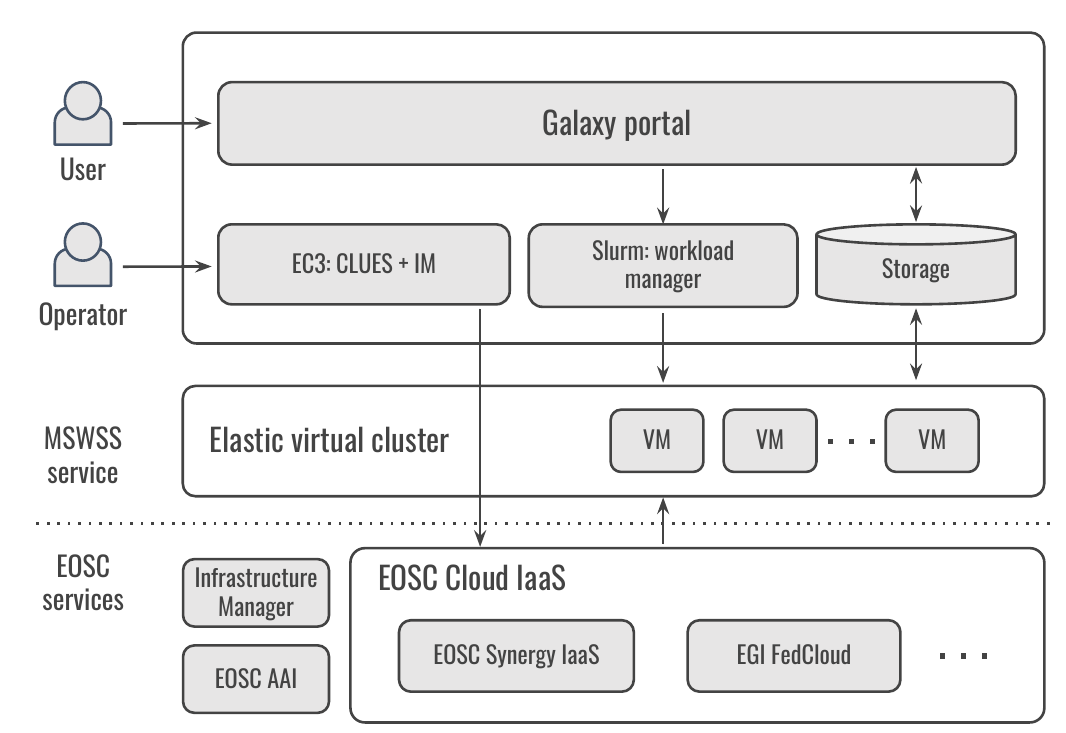}
	\end{center}
	\caption{\label{fig:MSWSS_arch}Architecture of the MSWSS service. }
\end{figure}

 Figure \ref{fig:MSWSS_arch} shows the architecture of the MSWSS service with the interaction of the service operator and users. The users interact with the service using Galaxy portal where they can manage their data and submit jobs to the elastic virtual cluster for processing. The output data are stored in within MSWSS service and can be downloaded for post-processing tasks.
 
 The service operator deploys the service using EC3 tool using the customised template. Once the MSWSS service is deployed the CLUES service monitors SLURM batch system and deploys new virtual worker nodes as needed and automatically re-configures the batch system. The deployment and configuration of worker nodes is performed using the Infrastructure Manager (IM) \cite{10.1007/s10723-014-9296-5} service. To speed-up the deployment process the worker nodes are instantiated from a snapshot of fully deployed worker node (golden image). This allows to decrease the start-up time from 21 to 5 minutes. It also helps to solve the issue with pending security updates with respect to the vanilla image and potential need to reboot the worker node for the updates to be applied properly. The golden image is maintained by the service operator in up-to-date state. The security is important also for the communication inside the virtual cluster. OpenVPN system is used to create secure connections inside the cluster and protect the data transfers. It also allows to span the virtual cluster over the resources from different Cloud providers.

\subsection{Data Storage}
Finally, in the field of data storage, we have also observed three different approaches: i) local storage; ii) external storage and iii) hybrid approach. The next subsections analyze the use cases of WORSICA, OpenEBench and UMSA to illustrate the three approaches.

\subsubsection{Local storage}
WORSICA uses the Dataverse application to manage the data produced by the service and disseminate it to the research community and the public in general. 

\begin{figure}[h!]
	\begin{center}
	\scalebox{.55}{\includegraphics[width = 1\linewidth]{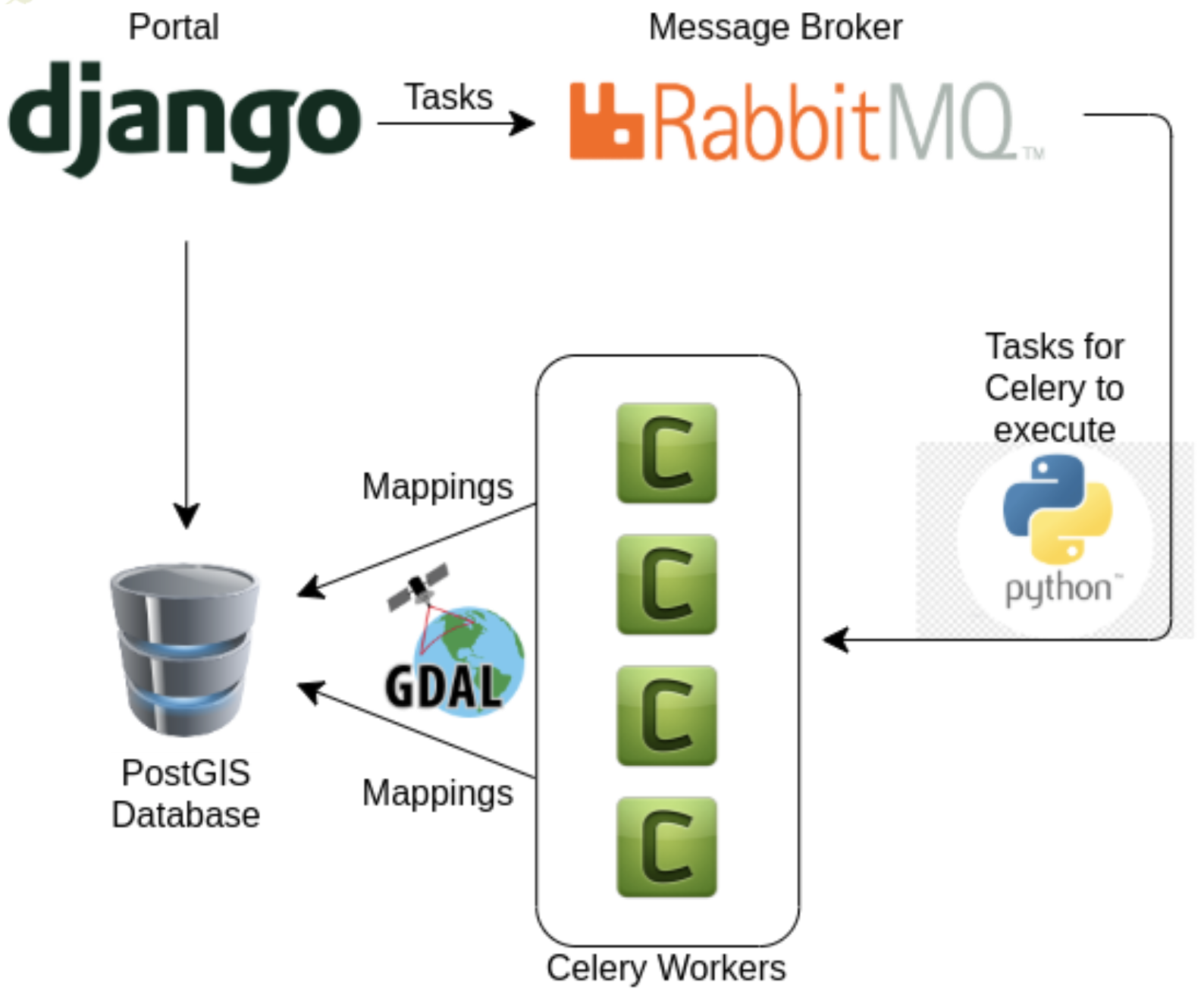}}
	\end{center}
	\caption{\label{fig:Worsica_arch1} First WORSICA's architecture for managing the data produced by the service.}
\end{figure}
The first version of WORSICA's data manager was developed as a local repository, using the architecture presented in Figure \ref{fig:Worsica_arch1}.
This approach raised several constraints to the adoption of a FAIR compliant data management paradigm, such as i) the lack of a unique global identifier for the datasets produced; ii) the data was not accessible and stored in multiple places internal to the service; iii) nonexistence of metadata for each dataset; and also iv) the access to the data did not follow the controlled vocabularies that apply to FAIR principles.  

\begin{figure}[h!]
	\begin{center}
	\scalebox{.65}{\includegraphics[width = 1\linewidth]{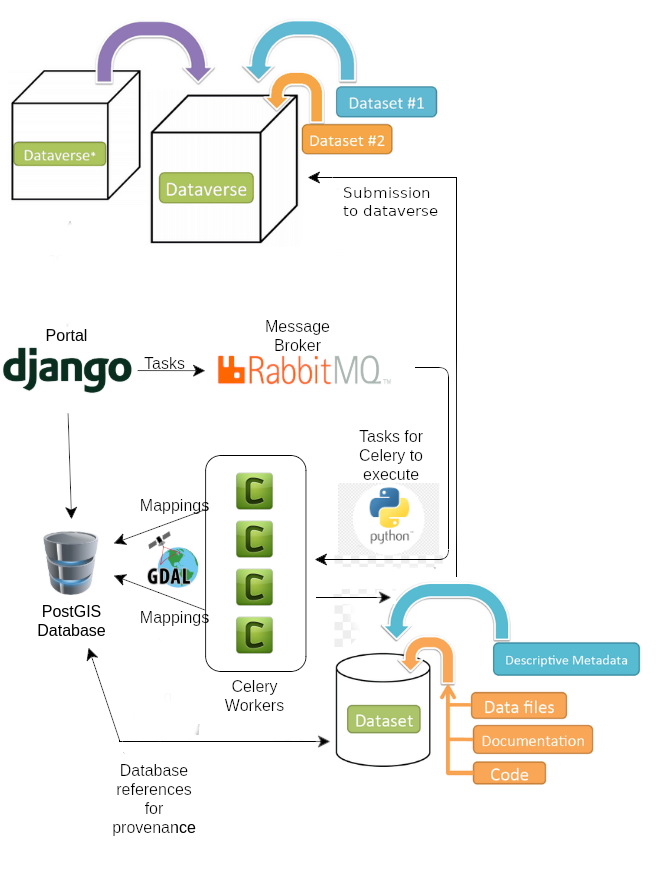}}
	\end{center}
	\caption{\label{fig:Worsica_arch2} Current architecture of the data manager implemented in the WORSICA service.}
\end{figure}
In order to surpass the previous oversights, a Dataverse REST API that allows running all necessary operations efficiently, like:
\begin{enumerate}
\item the ability to implement in any language only being dependent on the provided interface without any library requirements;
\item the capability to easily maintain the WORSICA code in parallel with Dataverse service updates;
\item Moreover, provide the features required to share sensitive data with the public.
\end{enumerate}

The current architecture of WORSICA's data manager evolved and can be seen in Figure \ref{fig:Worsica_arch2}.  
The WORSICA service is now working with Dataverse to automate the data availability and use services for and distribute credit to the data creator. Dataverse allows the creation of multiple virtual archives called Dataverse collections. Each Dataverse collection contains datasets, and each dataset contains descriptive metadata and data files. Therefore, this version of WORSICA enables datasets to be linked in Dataverse to the appropriate ontologies to increase interoperability and data FAIRness. Variable names can also be included in datasets metadata in the native language (Portuguese) and get Universal Resource Identifier (URI) for those entities in controlled vocabularies (e.g., in the case of WORSICA, a DOI - Digital Object Identifier is created). Furthermore, standardized metadata fields are available in Linked Open Data Cloud through standard machine-to-machine interfaces available in Dataverse.

\subsubsection{External storage}

OpenEBench makes use of B2SHARE for long-term availability and storage of scientific benchmarking datasets. The adoption of EUDAT’s technical standards, data models and policies helps OpenEbench to further enforce the  FAIR-compliant data management of the platform. One of the major capabilities gained through the integration with EUDAT is the minting of Digital Object Identifiers (DOIs) for the benchmarking data collections generated in OpenEBench.

A variety of dataset types are involved in the benchmarking workflows at OpenEBench. Those datasets cover the reference data used as gold standard data, the predictions submitted by participants as well as new datasets like the actual results scoring and ranging participants, provenance reports and metrics’ plots. In this way, a compact and human-readable set of data is ready to be referred to in scientific publications, promoting transparency, reproducibility and data reuse. Furthermore, EUDAT registries provide rich metadata fields for easing data discovery, so thus submitted collections could be annotated with cross-links to OpenEBench for further insights. Actually, OpenEBench is one of the EUDAT registered research communities benefiting from a particular extended metadata model and customized access rules. It facilitated a better integration with OpenEBench, implemented through a REST-based programmatic data publication workflow triggered from the platform Web GUI.

\begin{figure}[h]
	\begin{center}
	\scalebox{1.05}{\includegraphics[width = 1\linewidth]{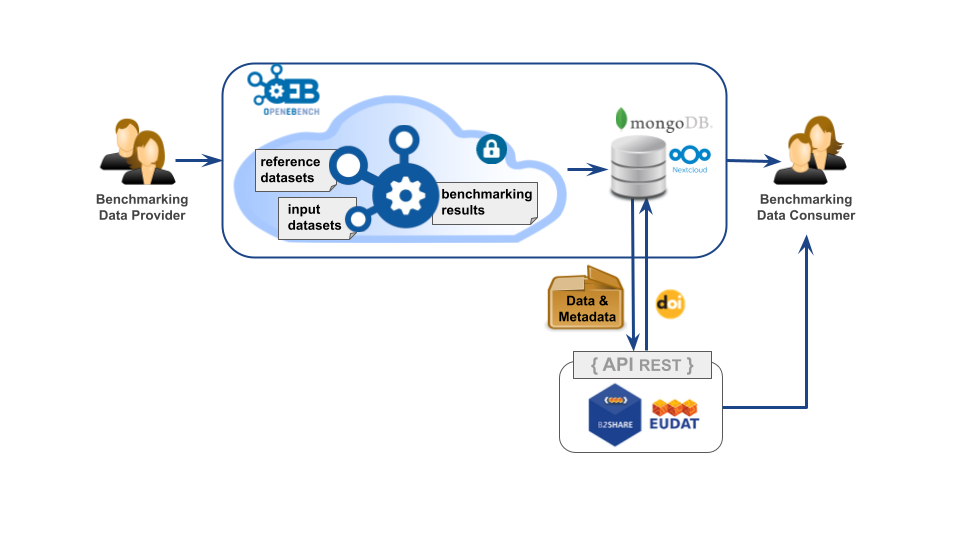}}
	\end{center}
	\caption{\label{fig:OEB_1} OpenEBench data flow. Users execute evaluation benchmarking workflows, which use and produce datasets. That data is locally stored in MongoDB (for metadata) and Nextcloud. It can be then submitted to B2SHARE, using their API REST, which will lead to minting a DOI. Then, it is mapped and saved to the OpenEBench database for cross-linking purposes. Published benchmarking data can be consumed using both B2SHARE and OpenBench platforms. }
\end{figure}

Within OpenEBench ecosystem, benchmarking data is accessible on the Web or via specific REST and GraphQL APIs. Nextcloud and MongoDB are the technologies used to store datasets and metadata respectively, always operated under a FAIR-compliant data governance plan that considers, for example, unique and accessible identifiers, document versioning, provenance preserving metadata, strict publication rules or a formal benchmarking data model. When a benchmarking events manager or developer participating in a given event is willing to publish its data outside the platform, they register their B2SHARE API token in OpenEBench and initiate the publication process to B2SHARE via the Web GUI:
\begin{itemize}
    \item OpenEBench composes the data collection with a specific metadata form, validates it, and programmatically submits both, the data and metadata, to the B2SHARE server.
    \item Over HTTPS and on behalf of the user, the platform implements the full EUDAT data publication workflow through the B2SHARE REST API. The outcome is a DOI associated with the new registry, which is captured and saved in OpenEBench to keep both systems cross-linked. Eventually, published benchmarking datasets can be consumed using both B2SHARE and OpenBench platforms.
\end{itemize}
The OpenEBench data flow can be seen in Figure \ref{fig:OEB_1}.  

\subsubsection{Hybrid approach}

UMSA leverages three different storage classes for different purposes.
First, data acquired by the instruments (mass spectrometers) are stored on a~traditional POSIX filesystem, 
implemented as RAID disk arrays and cluster of GPFS servers, re-exporting the volumes via NFS and CIFS to the clients.
This is a~technology with limited scaling (up to petabytes per filesystem), however, it is proven for decades and stable,
suitable for the primary experimental data that are irrecoverable otherwise.
This storage is also backed up weekly to a~remote site provided by the CZ national e-Infrastructure%
\footnote{\url{http://du.cesnet.cz/}}.

The UMSA service itself is based on Galaxy, which mounts the primary data filesystem (above) read-only,
and it exposes the datafiles to the users via its \emph{Data libraries}%
\footnote{\url{https://galaxyproject.org/data-libraries/}}, with appropriate access control,
and without the need to copy the large files.

Standard Galaxy setup requires a~shared filesystem mounted at both head and worker nodes
(alternate setups require lot of file transfers for each job, which we cannot afford).
It can be either a~single-tier storage, or it can serve as the first tier (cache) of 
object storage (the second tier).
We use the two-tier setup, with a~NFS-mounted SSD-only shared filesystem, with typical usage
up to dozens of TB only, fast enough not to slow down data processing with I/O,
and S3 object storage (provided by the national e-Infrastructure again), which scales up to many
PB easily.

\section{Related Work}
\label{sec:relatedwork}
This section describes other alternative thematic services and their differences with respect to the provisioning or performing model of the services in this article. Each thematic service has a dedicated paragraph to better analyze its case.

The WORSICA thematic service aims to study the evolution of coastal zones and water in inland areas for better coastal, estuarine, and natural resource management. Other services study these subjects, such as \cite{DEA,rs11242984, BISHOPTAYLOR2021112734, Dias}; however, the WORSICA service has the particularity of being freely accessible to the scientific community and performing simulations on-demand for the entire globe. Furthermore, the development of a service like WORSICA can only be achieved through the collaborative work of the EOSC network regarding SaaS infrastructures and products with federated access, described in Section 4, due to its need for a robust and scalable computational infrastructure to serve its users.

GCore thematic service leverages cloud computing power to enhance the processing resources available in a Space Ground Segment. The adoption of a dynamic cloud backend is a trend in different EO missions. Programs as Copernicus envisage the use of cloud computing to change the current on-premises processing approach to a on-the-cloud approach and offer the environments as a service. Entities as EUMETSAT (European Organization for the Exploitation of Meteorological Satellites) envisage a new design of a multimission processing infrastructure using the cloud advantages in order to allow extending the resources and made them available to several missions. GCore follows these approaches in order to extend the functionality with the use of cloud resources in order to break the bottleneck that a on premise classical system can obtain during the re-processing tasks associated to a particular mission for example. During this task a massive use of processing power is needed in the nominal platform  being able to affect to the nominal operations of the mission. The capacity of GCore to deploy additional processors on the cloud is used to reduce the impact of such a task. This approach can also be used for data archiving and cataloguing the products resulted from the processing in the cloud making them available to perform higher processing levels directly on cloud using as a service processors specifically defined published previously in a market place.

SAPS applies algorithms to estimate evapotranspiration (ET), a technique that has been widely used in several countries, and in areas subject to different weather conditions \cite{MU20111781}, \cite{Wan20156485}. The execution of these algorithms suppose high computational demands, both CPU and memory-intensive, and the archival of the output data generated consumes a substantial amount of storage resources \cite{GOODMAN2019103}. Recently, several software packages and libraries with implementations of ET estimation have been
made publicly available, such as the ILWIS software \cite{ABOUALI2013134} or the 'Water' procedure \cite{olmedo2017water}. However, those implementations have been developed as standalone artifacts made available for individual use by researchers, typically executed in their personal computers, that complicate considerably the sharing of processed data and might suppose limitations on the computing capacity locally available. A different approach is taken by the Google Earth Engine (GEE) platform \cite{googleearth} an initiative to facilitate the implementation and execution of scientific workflows
that consume satellite imagery as input. However, this service is under service conditions imposed by Google, that might not always fit the community needs \cite{PADARIAN201580}. Conversely, SAPS in combination with EOSC services and resources, follows an innovative approach based on the deployment of on-demand SAPS instances on top of federated resources. The service is highly configurable, allowing the selection of the algorithms used to estimate ET, and provides a standard output data and metadata that can be easily shared among researchers of the area.

The ScipionCloud service offers a ready-to-use infrastructure in the cloud to users that aim to process their CryoEM data. A similar approach is followed by Stion \cite{Stion}, a web application that provides on-demand access to GPU instances on AWS for biomedical researchers to process Cryo-EM data. This solution offers automatic deploying of instances (virtual machines) but does not integrate a batch system. It also sets up an auto-shutdown mechanism to power off instances after a certain period of time, which is risky and insufficient from our point of view. On the contrary plans to integrate ScipionCloud with EC3 guarantees a much better way to optimize the use of cloud resources than Stion's implementation. Besides, to use Stion it is mandatory to provide your own AWS account which might be a drawback for many users both in terms of complexity and cost. Another approach found in the CryoEM world is the work done by Cianfrocco's lab at the University of Michigan, which offers several tools in this area. The first interesting tool \citep{Cianfrocco2018} is the integration of a 'AWS batch system' in one of the most popular software packages used for CryoEM processing; Relion. This batch system allows to send Relion commands to AWS instances, which includes deploying, running and shutting down the instance. This approach was only integrated in an old version of Relion which might imply that it was not a big success in the CryoEM community. The second tool is COSMIC \citep{Cianfrocco2017}, a freely available web platform for submitting cryo-EM jobs through the cloud to Comet, a cluster at the San Diego Supercomputer Center. Although the use of COSMIC has no cost access and this tool is available for everybody users might prefer to work on their own server instead of preparing their processing workflows to be sent step by step to a cluster.

Regarding the use of external storages for sustainability, OpenEBench is not the only platform designed around FAIR principles that ensures the availability of its data and metadata through EUDAT services or similar data infrastructures. WorkflowHub \cite{WorkflowHub} is applying a similar strategy for publishing scientific workflows, wrapping them into enriched Research Object Crates (RO-Crate 1.1 specification), which include data resources, semantic annotations and all additional information that guarantees the workflow is reusable and reproducible. In this case, Zenodo is the infrastructure minting the DOI, and the publication process follows DataCite \citep{DataCite} guidelines, one of the broadest cross-domain metadata standards available. Among EOSC resources, other platforms like ROHub \citep{ROHub} propose the use of EUDAT services as an internal storage solution. ROHub is a platform that enables the management, sharing and preservation of research data as Research Objects. It is integrating B2DROP \citep{B2DROP} as the underlying technology for the researcher's personal storage space and B2SHARE for DOI minting and sharing.

Software needed for LAGO simulations is managed by ARTI\,\cite{Asorey2018}, a data-intensive and highly-complex framework designed to calculate the expected flux of signals in any site around the World and under realistic geomagnetic, atmospheric and detector conditions. As the simulated phenomena, i.e., the interaction of cosmic rays with the atmosphere and the detector response to the resulting flux secondary particles, is, essentially, a sequence of stochastic processes, the simulation performed need to integrate the flux over long periods to reduce the impact of statistical fluctuations. Such large times, from several hours to days and even years for some applications such as volcano risk assessments, require the usage of large computing facilities and storage. Suitable simulations will typically spend tens of weeks in current CPUs and they will output several TB of data. Examples of that are the results presented in Subsection\,\ref{LAGO:fixed}, and recently in \cite{LAGO_WSC21,LAGO-icrc2021}. Moreover, all these data needs to be properly identified, catalogued and curated to accomplish the FAIR principles. Although previous attempts were made to adapt ARTI to its use in distributed high-performance computing infrastructures\,\cite{Asorey2016} and for the adoption of data curation standards\,\cite{Rodriguez2015}, it was only thanks to the development of OnedataSim\,\cite{LAGO_WSC21} within the LAGO TS that it was possible to achieve sufficiently long integration times while complying FAIR principles.

SDS-WAS provide, a part of a huge quantity of materials on dust storms forecast, services of data storage, download and data analysis based on EOSC cloud services (B2SAFE, B2HANDLE). As far as we know there aren't comparable services related to dust storms, in terms of collecting a bunch of numerical models outputs, observational data (in situ and satellite), providing derived graphical (plots) and numerical (skill scores) products with an interactive dashboard application. Those services are related to the areas of Northern Africa, Middle-East and Europe.
We considered as "competitors" the other SDS-WAS regional nodes (Asian and Panamerican), but also other services (not only provided by Research Institutions, but also by companies) related not directly to dust, but to air quality forecast or similar. The aim is to see how they approach the information and solve user's needs. While the Asian\footnote{\url{http://www.asdf-bj.net/}} and Panamerican\footnote{\url{http://sds-was.cimh.edu.bb/}} SDS-WAS nodes are still in a preliminary stage, as they provide a limited set of services, there are some others with a good feedback. We analysed strengths and weaknesses of products visual presentation of following services: METOFFICE\footnote{\url{https://www.metoffice.gov.uk/}}, Breezometer \footnote{\url{https://breezometer.com/}}, WINDY\footnote{\url{https://www.windy.com/}}, Plume Labs\footnote{\url{https://air.plumelabs.com/}}. Although most of them have an attractive design, they only offer visual products of their data, not numerical binaries, and the amount of data they provides seems to be not big as what our SDS-WAS service does. Finally, some of them have advertisements and banners that affects significantly the user experience.

UMSA is dedicated to storage and processing mass-spectrometry data, focusing on GC-MS and untargeted analysis of low-abundant compounds. Numerous tools to process MS data exist, ranging from proprietary software by laboratory equipment vendors, third party commercial software, and open source tools of widely varying quality and maturity.
Several sub-area specific reviews exist~\cite{msprot,YI201617}.
Being based on Galaxy, UMSA can leverage from dozens of mass-spec related tools published by the community.
There is ongoing effort to provide Galaxy-based environments focusing on specific MS applications, e.g.~\cite{wf4mb}.
According to our knowledge, there is no such effort matching the UMSA purpose. Another approach can be seen with GNPS~\cite{gnps}, which is a~web-based computational environment and data treatment environment,
centered around\emph{molecular networking}---visual display of chemical spaces and relationships among compounds.
Vast majority of the methods leverage on MS$^2$ data, which is not the focus of UMSA.
On the other hand, publication and storage of the results of the analyses is fairly well organized around the MassBank project~\cite{massbank}.

The MSWSS is a cloud-based service capable to exploit the resources from the EOSC cloud infrastructure and FAIR data repositories. There are several approaches for modelling water supply systems offering various features. Integrated Tool for Water Supply Systems Management \cite{perez_2021} puts together QGIS database, Epanet hydraulic model, and Google Maps. A web-based EPANET model catalogue and execution environment \cite{bayer_2021} focuses on model sharing and a model viewing. Another example of integration GIS and the hydraulic model is a tool for an integrated water and wastewater management system in municipal enterprises \cite{kruszynski_2020}. According to our knowledge these services do not provide cloud-based elastic computational back-end and they do not implement FAIR principles for data sharing.

O3as has been inspired by the need of having the large plethora of climate model data available in a consistent manner. Because of the ozone assessment\footnote{\url{https://csl.noaa.gov/assessments/ozone/index.html} [2022-02-09]} that culminates in a publication every four years \cite{WMO_2018} scientists need to have a quick way on looking at ozone data from climate data to estimate the time when the amount of ozone in the stratosphere has reached a level pre-ozone hole. Often the models have to be collected, the analysis program redone and the values recalculated \cite{dhomse_2018, keeble_2021}. The O3as thematic service is very modular and can therefore be extended so that not only zonally  (i.e. across longitudes) averaged values of ozone data will be shown, but also trends and ozone return rates for different positions on the Earth. Also atmospheric trace gases other than ozone could in the future be included into the analysis. 

\section{Conclusions}
\label{sec:conclusions}

EOSC-SYNERGY is building capacities in EOSC through the development of ten data-intensive thematic services oriented to four different scientific disciplines. The adaptation, improvement and quality assessment of those services on a federated data infrastructure strongly aligns with the objectives of EOSC \cite{eoscdraft}. A key factor for the success of EOSC \cite{doi/10.2777/870770} is performance, i.e, how EOSC as an ecosystem operates and how the resources are used and acknowledged by the users. All the services consume services from the EOSC catalogue, which will provide feedback on the usability and relevance of the model.

However, the characterisation of new applications to join is key to evaluate the rightmost services to be used. This is why the EOSC-Synergy project is developing best practices and experiences to promote EOSC adoption by the research communities by expanding and building knowledge on common interfaces, standards and best practices. This paper presented an application modelling proposal, based on the previous analysis of gaps and bottlenecks performed by ten different services from four different disciplines. 

The ten thematic services have been analysed concerning four dimensions (authentication and authorisation; resource management and offering model; workload management including containerisation; and data storage and preservation), identifying gaps and bottlenecks. Those requirements are common to many other scientific services. We selected ten services from the EOSC Marketplace to address these requirements. In a nutshell, AAI solutions such EGI Checkin, eduTEAMS, D2Access and life-science AAI are mature enough to provide a coherent authentication model for a whole application. Applications that require a dynamic backend or on-demand deployment found Infrastructure Manager (IM) and Elastic Compute Clusters in the Cloud (EC3) as reasonable solutions to describe their infrastructure as code and deploy resources according to their workload. Depending on the workload type, job management is driven by SLURM batch queues or Kubernetes services. Preservation of data is obtained through EGI DataHub or EUDAT's B2SAFE and B2SHARE storage services, registering persistent identifiers through B2Handle. Finally, services needing local storage used Dataverse as an OAI-PMH on-premise storage.

The impact of EOSC in the thematic services of EOSC Synergy is mainly composed of three main facts. Firstly, the capacity expansion through the federation of computing, storage, and data resources aligned with the EOSC and FAIR policies and practices. Secondly, software and service quality evaluation of the thematic services is critical to improve robustness and reliance and therefore increase user's experience. EOSC-SYNERGY also focuses on transverse training to facilitate the adoption of technologies and the use of thematic services. Finally the cross-fertilization between different thematic areas has allowed the collaboration between thematic services to take advantage of the developments, solutions, experiences and best practices on authentication and authorization, data storage, resource and workload management.

\section*{Acknowledgements}
This work was supported by the European Union’s Horizon 2020 research and innovation programme under grant agreement No 857647, EOSC-synergy, European Open Science Cloud - Expanding Capacities by building Capabilities. Moreover, this work is partially funded by grant No 2015/24461-2, S\~{a}o Paulo Research Foundation (FAPESP). Francisco Brasileiro is a CNPq/Brazil researcher (grant 308027/2020-5).

\bibliographystyle{elsarticle-num-names} 

\bibliography{main.bib}

\begin{thebibliography}{84}
\expandafter\ifx\csname natexlab\endcsname\relax\def\natexlab#1{#1}\fi
\providecommand{\url}[1]{\texttt{#1}}
\providecommand{\href}[2]{#2}
\providecommand{\path}[1]{#1}
\providecommand{\DOIprefix}{doi:}
\providecommand{\ArXivprefix}{arXiv:}
\providecommand{\URLprefix}{URL: }
\providecommand{\Pubmedprefix}{pmid:}
\providecommand{\doi}[1]{\href{http://dx.doi.org/#1}{\path{#1}}}
\providecommand{\Pubmed}[1]{\href{pmid:#1}{\path{#1}}}
\providecommand{\bibinfo}[2]{#2}
\ifx\xfnm\relax \def\xfnm[#1]{\unskip,\space#1}\fi
\bibitem[{Staff(2021)}]{ieee20212021}
\bibinfo{author}{I.~Staff}, \bibinfo{title}{2021 IEEE 17th International
  Conference on EScience (eScience)}, \bibinfo{publisher}{IEEE},
  \bibinfo{year}{2021}. \URLprefix
  \url{https://books.google.es/books?id=o924zgEACAAJ}.
\bibitem[{Foster(2016)}]{opensciencedef}
\bibinfo{author}{Foster}, \bibinfo{title}{Open science definition},
  \bibinfo{howpublished}{\url{https://www.fosteropenscience.eu/foster-taxonomy/open-science-definition}},
  \bibinfo{year}{2016}.
\bibitem[{Commission(2016)}]{einfrastructures}
\bibinfo{author}{E.~Commission}, \bibinfo{title}{e-infrastructures definition},
  \bibinfo{howpublished}{\url{https://ec.europa.eu/digital-single-market/en/e-infrastructures}},
  \bibinfo{year}{2016}.
\bibitem[{Blanquer et~al.(2012)Blanquer, Brasche, and
  Lezzi}]{blanquer2012requirements}
\bibinfo{author}{I.~Blanquer}, \bibinfo{author}{G.~Brasche},
  \bibinfo{author}{D.~Lezzi},
\newblock \bibinfo{title}{Requirements of scientific applications in cloud
  offerings},
\newblock in: \bibinfo{booktitle}{Proceedings of the 2012 Sixth Iberian Grid
  Infrastructure Conference, IBERGRID}, volume~\bibinfo{volume}{12},
  \bibinfo{year}{2012}, pp. \bibinfo{pages}{173--182}.
\bibitem[{Group(2020)}]{fair_data_maturity_model_working_group_2020_3909563}
\bibinfo{author}{F.~D. M. M.~W. Group}, \bibinfo{title}{{FAIR Data Maturity
  Model. Specification and Guidelines}}, \bibinfo{year}{2020}. \URLprefix
  \url{https://doi.org/10.15497/rda00050}. \DOIprefix\doi{10.15497/rda00050}.
\bibitem[{Commission(2020)}]{eosc}
\bibinfo{author}{E.~Commission}, \bibinfo{title}{Eosc european partnership
  proposal}, \bibinfo{howpublished}{\url{https://bit.ly/2RKYmak}},
  \bibinfo{year}{2020}.
\bibitem[{Synergy(2020)}]{eoscsynergy}
\bibinfo{author}{E.~Synergy}, \bibinfo{title}{Eosc synergy portal},
  \bibinfo{howpublished}{\url{https://www.eosc-synergy.eu/}},
  \bibinfo{year}{2020}.
\bibitem[{Enhance and Future(2021)}]{eosc_catalog}
\bibinfo{author}{E.~Enhance}, \bibinfo{author}{E.~Future}, \bibinfo{title}{Eosc
  portal catalogue \& marketplace},
  \bibinfo{howpublished}{\url{https://marketplace.eosc-portal.eu/services}},
  \bibinfo{year}{2021}.
\bibitem[{WORSICA(2021{\natexlab{a}})}]{worsica_service1}
\bibinfo{author}{WORSICA}, \bibinfo{title}{{LNEC Portal - Water Monitoring
  Sentinel Cloud Platform}},
  \bibinfo{howpublished}{\url{http://worsica.lnec.pt}},
  \bibinfo{year}{2021}{\natexlab{a}}.
\bibitem[{WORSICA(2021{\natexlab{b}})}]{worsica_service2}
\bibinfo{author}{WORSICA}, \bibinfo{title}{{Water Monitoring Sentinel Cloud
  Platform}}, \bibinfo{howpublished}{\url{https://worsica.incd.pt/index/}},
  \bibinfo{year}{2021}{\natexlab{b}}.
\bibitem[{Cunha et~al.(2020)Cunha, Pereira, Pereira, Rufino, Galvão, Valente,
  and Brasileiro}]{CUNHA2020104341}
\bibinfo{author}{J.~Cunha}, \bibinfo{author}{T.~E. Pereira},
  \bibinfo{author}{E.~Pereira}, \bibinfo{author}{I.~Rufino},
  \bibinfo{author}{C.~Galvão}, \bibinfo{author}{F.~Valente},
  \bibinfo{author}{F.~Brasileiro},
\newblock \bibinfo{title}{A high-throughput shared service to estimate
  evapotranspiration using landsat imagery},
\newblock \bibinfo{journal}{Computers \& Geosciences} \bibinfo{volume}{134}
  (\bibinfo{year}{2020}) \bibinfo{pages}{104341}. \URLprefix
  \url{https://www.sciencedirect.com/science/article/pii/S0098300419302961}.
  \DOIprefix\doi{https://doi.org/10.1016/j.cageo.2019.104341}.
\bibitem[{Zapata(2019)}]{gcore_poster}
\bibinfo{author}{M.~Zapata}, \bibinfo{title}{{SMOS Fast Reprocessing Platform
  at ESAC}}, \bibinfo{howpublished}{\url{shorturl.at/lwyQ2}},
  \bibinfo{year}{2019}. \bibinfo{note}{Living planet symposium, 13-17 May 2019,
  Milan, Italy}.
\bibitem[{Rodriguez(2018)}]{gcore_esa}
\bibinfo{author}{M.~Rodriguez},
  \bibinfo{howpublished}{\url{shorturl.at/dfoIQ}}, \bibinfo{year}{2018}.
\bibitem[{GCore(2022)}]{gcore_service}
\bibinfo{author}{GCore}, \bibinfo{title}{{GCore overview at EOSC Synergy}},
  \bibinfo{howpublished}{\url{https://www.eosc-synergy.eu/thematic-services/g-core/}},
  \bibinfo{year}{2022}.
\bibitem[{{de la Rosa-Trevín} et~al.(2016){de la Rosa-Trevín}, Quintana, {del
  Cano}, Zaldívar, Foche, Gutiérrez, Gómez-Blanco, Burguet-Castell,
  Cuenca-Alba, Abrishami, Vargas, Otón, Sharov, Vilas, Navas, Conesa, Kazemi,
  Marabini, Sorzano, and Carazo}]{DELAROSATREVIN201693}
\bibinfo{author}{J.~{de la Rosa-Trevín}}, \bibinfo{author}{A.~Quintana},
  \bibinfo{author}{L.~{del Cano}}, \bibinfo{author}{A.~Zaldívar},
  \bibinfo{author}{I.~Foche}, \bibinfo{author}{J.~Gutiérrez},
  \bibinfo{author}{J.~Gómez-Blanco}, \bibinfo{author}{J.~Burguet-Castell},
  \bibinfo{author}{J.~Cuenca-Alba}, \bibinfo{author}{V.~Abrishami},
  \bibinfo{author}{J.~Vargas}, \bibinfo{author}{J.~Otón},
  \bibinfo{author}{G.~Sharov}, \bibinfo{author}{J.~Vilas},
  \bibinfo{author}{J.~Navas}, \bibinfo{author}{P.~Conesa},
  \bibinfo{author}{M.~Kazemi}, \bibinfo{author}{R.~Marabini},
  \bibinfo{author}{C.~Sorzano}, \bibinfo{author}{J.~Carazo},
\newblock \bibinfo{title}{Scipion: A software framework toward integration,
  reproducibility and validation in 3d electron microscopy},
\newblock \bibinfo{journal}{Journal of Structural Biology}
  \bibinfo{volume}{195} (\bibinfo{year}{2016}) \bibinfo{pages}{93--99}.
  \URLprefix
  \url{https://www.sciencedirect.com/science/article/pii/S104784771630079X}.
  \DOIprefix\doi{https://doi.org/10.1016/j.jsb.2016.04.010}.
\bibitem[{Ope(2017)}]{OpenEBench}
\bibinfo{title}{Lessons learned: Recommendations for establishing critical
  periodic scientific benchmarking},
\newblock \bibinfo{journal}{bioRxiv}  (\bibinfo{year}{2017}). \URLprefix
  \url{https://www.biorxiv.org/content/early/2017/08/31/181677}.
  \DOIprefix\doi{10.1101/181677}.
\bibitem[{ELIXIR(2021)}]{elixir}
\bibinfo{author}{ELIXIR}, \bibinfo{title}{Elixir european intergovernmental
  organisation}, \bibinfo{howpublished}{\url{https://elixir-europe.org/}},
  \bibinfo{year}{2021}.
\bibitem[{Sidelnik et~al.(2017)Sidelnik, Asorey, Collaboration
  et~al.}]{lagoproject}
\bibinfo{author}{I.~Sidelnik}, \bibinfo{author}{H.~Asorey},
  \bibinfo{author}{L.~Collaboration}, et~al.,
\newblock \bibinfo{title}{{LAGO: The Latin American giant observatory}},
\newblock \bibinfo{journal}{Nuclear Instruments and Methods in Physics Research
  Section A: Accelerators, Spectrometers, Detectors and Associated Equipment}
  \bibinfo{volume}{876} (\bibinfo{year}{2017}) \bibinfo{pages}{173--175}.
  \DOIprefix\doi{10.1016/j.nima.2017.02.069}.
\bibitem[{{LAGO Collaboration}(2021)}]{lagoprojectweb}
\bibinfo{author}{{LAGO Collaboration}}, \bibinfo{title}{{Latin American Giant
  Observatory}}, \bibinfo{howpublished}{\url{http://lagoproject.net}},
  \bibinfo{year}{Accessed, December 2021}.
\bibitem[{Rubio-Montero et~al.(2021)Rubio-Montero, Pag{\'a}n-Muñoz,
  Mayo-Garc{\'i}a, Pardo-D{\'i}az, Sildelnik, and Asorey}]{LAGO_WSC21}
\bibinfo{author}{A.~J. Rubio-Montero}, \bibinfo{author}{R.~Pag{\'a}n-Muñoz},
  \bibinfo{author}{R.~Mayo-Garc{\'i}a}, \bibinfo{author}{A.~Pardo-D{\'i}az},
  \bibinfo{author}{I.~Sildelnik}, \bibinfo{author}{H.~Asorey},
\newblock \bibinfo{title}{{A Novel Cloud-based Framework for Standardized
  Simulations in the Latin American Giant Observatory (LAGO)}},
\newblock in: \bibinfo{booktitle}{2021 Winter Simulation Conference (WSC)},
  \bibinfo{publisher}{IEEE}, \bibinfo{address}{Phoenix, USA.},
  \bibinfo{year}{13-17 Dec. 2021}. \bibinfo{note}{In Print.}
\bibitem[{{Basart} et~al.(2019){Basart}, {Nickovic}, {Terradellas}, {Cuevas},
  {P{\'e}rez Garc{\'\i}a-Pando}, {Garc{\'\i}a-Castrillo}, {Werner}, and
  {Benincasa}}]{sdswas}
\bibinfo{author}{S.~{Basart}}, \bibinfo{author}{S.~{Nickovic}},
  \bibinfo{author}{E.~{Terradellas}}, \bibinfo{author}{E.~{Cuevas}},
  \bibinfo{author}{C.~{P{\'e}rez Garc{\'\i}a-Pando}},
  \bibinfo{author}{G.~{Garc{\'\i}a-Castrillo}}, \bibinfo{author}{E.~{Werner}},
  \bibinfo{author}{F.~{Benincasa}},
\newblock \bibinfo{title}{{The WMO SDS-WAS Regional Center for Northern Africa,
  Middle East and Europe}},
\newblock in: \bibinfo{booktitle}{E3S Web of Conferences},
  volume~\bibinfo{volume}{99} of \textit{\bibinfo{series}{E3S Web of
  Conferences}}, \bibinfo{year}{2019}, p. \bibinfo{pages}{04008}.
  \DOIprefix\doi{10.1051/e3sconf/20199904008}.
\bibitem[{{Basart} et~al.(2015){Basart}, {Terradellas}, {Cuevas}, {Jorba},
  {Benincasa}, and {Baldasano}}]{bdfc}
\bibinfo{author}{S.~{Basart}}, \bibinfo{author}{E.~{Terradellas}},
  \bibinfo{author}{E.~{Cuevas}}, \bibinfo{author}{O.~{Jorba}},
  \bibinfo{author}{F.~{Benincasa}}, \bibinfo{author}{J.~M. {Baldasano}},
\newblock \bibinfo{title}{{The Barcelona Dust Forecast Center: The first WMO
  regional meteorological center specialized on atmospheric sand and dust
  forecast}},
\newblock in: \bibinfo{booktitle}{EGU General Assembly Conference Abstracts},
  EGU General Assembly Conference Abstracts, \bibinfo{year}{2015}, p.
  \bibinfo{pages}{13309}.
\bibitem[{UMSA(2022)}]{umsa_service}
\bibinfo{author}{UMSA}, \bibinfo{title}{{UMSA overview at EOSC Synergy}},
  \bibinfo{howpublished}{\url{https://www.eosc-synergy.eu/thematic-services/umsa/}},
  \bibinfo{year}{2022}.
\bibitem[{MSWSS(2021)}]{mswss_service}
\bibinfo{author}{MSWSS}, \bibinfo{title}{Modelling service for water supply
  systems}, \bibinfo{howpublished}{\url{https://mswss.ui.savba.sk:8443}},
  \bibinfo{year}{2021}.
\bibitem[{O3AS(2022)}]{o3as_service}
\bibinfo{author}{O3AS}, \bibinfo{title}{{O3AS Portal - Ozone Assessment Cloud
  Platform}}, \bibinfo{howpublished}{\url{https://o3as.data.kit.edu}},
  \bibinfo{year}{2022}.
\bibitem[{Yoo et~al.(2003)Yoo, Jette, and Grondona}]{slurm}
\bibinfo{author}{A.~B. Yoo}, \bibinfo{author}{M.~A. Jette},
  \bibinfo{author}{M.~Grondona},
\newblock \bibinfo{title}{Slurm: Simple linux utility for resource management},
\newblock in: \bibinfo{editor}{D.~Feitelson}, \bibinfo{editor}{L.~Rudolph},
  \bibinfo{editor}{U.~Schwiegelshohn} (Eds.), \bibinfo{booktitle}{Job
  Scheduling Strategies for Parallel Processing}, \bibinfo{publisher}{Springer
  Berlin Heidelberg}, \bibinfo{address}{Berlin, Heidelberg},
  \bibinfo{year}{2003}, pp. \bibinfo{pages}{44--60}.
\bibitem[{Goecks et~al.(????)Goecks, Nekrutenko, and Taylor}]{goecks11t}
\bibinfo{author}{J.~Goecks}, \bibinfo{author}{A.~Nekrutenko},
  \bibinfo{author}{J.~Taylor},
\newblock \bibinfo{title}{T.(2010) galaxy: a comprehensive approach for
  supporting accessible, reproducible, and transparent computational research
  in the life sciences},
\newblock \bibinfo{journal}{Genome Biol} \bibinfo{volume}{11} (????)
  \bibinfo{pages}{R86}.
\bibitem[{King(2007)}]{King07}
\bibinfo{author}{G.~King},
\newblock \bibinfo{title}{An introduction to the dataverse network as an
  infrastructure for data sharing},
\newblock \bibinfo{journal}{Sociological Methods and Research}
  \bibinfo{volume}{36} (\bibinfo{year}{2007})
  \bibinfo{pages}{173{\textendash}199}.
\bibitem[{Viljoen et~al.(2016)Viljoen, Łukasz Dutka, Kryza, and
  Chen}]{VILJOEN2016148}
\bibinfo{author}{M.~Viljoen}, \bibinfo{author}{Łukasz Dutka},
  \bibinfo{author}{B.~Kryza}, \bibinfo{author}{Y.~Chen},
\newblock \bibinfo{title}{Towards european open science commons: The egi open
  data platform and the egi datahub},
\newblock \bibinfo{journal}{Procedia Computer Science} \bibinfo{volume}{97}
  (\bibinfo{year}{2016}) \bibinfo{pages}{148--152}. \URLprefix
  \url{https://www.sciencedirect.com/science/article/pii/S187705091632110X}.
  \DOIprefix\doi{https://doi.org/10.1016/j.procs.2016.08.294},
  \bibinfo{note}{2nd International Conference on Cloud Forward: From
  Distributed to Complete Computing}.
\bibitem[{Lecarpentier et~al.(2013)Lecarpentier, Wittenburg, Elbers, Michelini,
  Kanso, Coveney, and Baxter}]{lecarpentier2013eudat}
\bibinfo{author}{D.~Lecarpentier}, \bibinfo{author}{P.~Wittenburg},
  \bibinfo{author}{W.~Elbers}, \bibinfo{author}{A.~Michelini},
  \bibinfo{author}{R.~Kanso}, \bibinfo{author}{P.~Coveney},
  \bibinfo{author}{R.~Baxter},
\newblock \bibinfo{title}{Eudat: a new cross-disciplinary data infrastructure
  for science},
\newblock \bibinfo{journal}{International Journal of Digital Curation}
  \bibinfo{volume}{8} (\bibinfo{year}{2013}) \bibinfo{pages}{279--287}.
\bibitem[{EGI(2021{\natexlab{a}})}]{EGICLOUD}
\bibinfo{author}{EGI}, \bibinfo{title}{Egi cloud compute service},
  \bibinfo{howpublished}{\url{https://www.egi.eu/services/cloud-compute/}},
  \bibinfo{year}{2021}{\natexlab{a}}.
\bibitem[{EGI(2021{\natexlab{b}})}]{EGIHTC}
\bibinfo{author}{EGI}, \bibinfo{title}{Egi high-throughput compute},
  \bibinfo{howpublished}{\url{https://www.egi.eu/services/high-throughput-compute/}},
  \bibinfo{year}{2021}{\natexlab{b}}.
\bibitem[{EGI(2021{\natexlab{c}})}]{EGIWM}
\bibinfo{author}{EGI}, \bibinfo{title}{Egi workload manager},
  \bibinfo{howpublished}{\url{https://www.egi.eu/services/workload-manager/}},
  \bibinfo{year}{2021}{\natexlab{c}}.
\bibitem[{EGI(2021{\natexlab{d}})}]{EGIDATAHUB}
\bibinfo{author}{EGI}, \bibinfo{title}{Egi data hub},
  \bibinfo{howpublished}{\url{https://www.egi.eu/services/datahub/}},
  \bibinfo{year}{2021}{\natexlab{d}}.
\bibitem[{EUDAT(2021)}]{B2SAFE}
\bibinfo{author}{EUDAT}, \bibinfo{title}{B2safe, keep research data safe via
  data management policies},
  \bibinfo{howpublished}{\url{https://sp.eudat.eu/catalog/resources/5d81cb5b-3640-4430-b46e-fc652e06a4db}},
  \bibinfo{year}{2021}.
\bibitem[{Caballer et~al.(2015)Caballer, Blanquer, Molt{\'{o}}, and
  de~Alfonso}]{Caballer2015}
\bibinfo{author}{M.~Caballer}, \bibinfo{author}{I.~Blanquer},
  \bibinfo{author}{G.~Molt{\'{o}}}, \bibinfo{author}{C.~de~Alfonso},
\newblock \bibinfo{title}{{Dynamic Management of Virtual Infrastructures}},
\newblock \bibinfo{journal}{Journal of Grid Computing} \bibinfo{volume}{13}
  (\bibinfo{year}{2015}) \bibinfo{pages}{53--70}.
  \DOIprefix\doi{10.1007/s10723-014-9296-5}.
\bibitem[{EGI(2021)}]{DYNDNS}
\bibinfo{author}{EGI}, \bibinfo{title}{Dynamic dns for vms in egi cloud},
  \bibinfo{howpublished}{\url{https://docs.egi.eu/users/cloud-compute/dynamic-dns/}},
  \bibinfo{year}{2021}.
\bibitem[{Tran(2021)}]{EGIFedCloudCli}
\bibinfo{author}{V.~Tran}, \bibinfo{title}{Fedcloud client documentation},
  \bibinfo{howpublished}{\url{https://fedcloudclient.fedcloud.eu/}},
  \bibinfo{year}{2021}.
\bibitem[{EUDAT(2021)}]{B2FIND}
\bibinfo{author}{EUDAT}, \bibinfo{title}{B2find official webpage, find research
  data, research data portal},
  \bibinfo{howpublished}{\url{https://sp.eudat.eu/catalog/resources/33bc21d5-f53d-4eed-9a15-56f98f5c7f69}},
  \bibinfo{year}{2021}.
\bibitem[{Calatrava et~al.(2016)Calatrava, Romero, Moltó, Caballer, and
  Alonso}]{CALATRAVA201613}
\bibinfo{author}{A.~Calatrava}, \bibinfo{author}{E.~Romero},
  \bibinfo{author}{G.~Moltó}, \bibinfo{author}{M.~Caballer},
  \bibinfo{author}{J.~M. Alonso},
\newblock \bibinfo{title}{Self-managed cost-efficient virtual elastic clusters
  on hybrid cloud infrastructures},
\newblock \bibinfo{journal}{Future Generation Computer Systems}
  \bibinfo{volume}{61} (\bibinfo{year}{2016}) \bibinfo{pages}{13--25}.
  \URLprefix
  \url{https://www.sciencedirect.com/science/article/pii/S0167739X16300024}.
  \DOIprefix\doi{https://doi.org/10.1016/j.future.2016.01.018}.
\bibitem[{EGI(2021)}]{EGICHECKIN}
\bibinfo{author}{EGI}, \bibinfo{title}{Egi check-in},
  \bibinfo{howpublished}{\url{https://www.egi.eu/services/check-in/}},
  \bibinfo{year}{2021}.
\bibitem[{EUDAT(2021)}]{B2ACCESS}
\bibinfo{author}{EUDAT}, \bibinfo{title}{Official webpage b2access identity \&
  authorisation},
  \bibinfo{howpublished}{\url{https://sp.eudat.eu/catalog/resources/d04af0f5-2253-4ee4-8181-3a5a961ccd49}},
  \bibinfo{year}{2021}.
\bibitem[{GEANT(2021)}]{EDUTEAMS}
\bibinfo{author}{GEANT}, \bibinfo{title}{eduteams web site/},
  \bibinfo{howpublished}{\url{https://eduteams.org/}}, \bibinfo{year}{2021}.
\bibitem[{Linden~M(2018)}]{ELIXIRAAI}
\bibinfo{author}{L.~I. e.~a. Linden~M, Procházka~M},
\newblock \bibinfo{title}{Common elixir service for researcher authentication
  and authorisation},
\newblock \bibinfo{journal}{F1000Research} \bibinfo{volume}{7}
  (\bibinfo{year}{2018}).
  \DOIprefix\doi{https://doi.org/10.12688/f1000research.15161.1}.
\bibitem[{Binz et~al.(2014)Binz, Breitenb{\"u}cher, Kopp, and
  Leymann}]{Binz2014}
\bibinfo{author}{T.~Binz}, \bibinfo{author}{U.~Breitenb{\"u}cher},
  \bibinfo{author}{O.~Kopp}, \bibinfo{author}{F.~Leymann},
  \bibinfo{title}{TOSCA: Portable Automated Deployment and Management of Cloud
  Applications}, \bibinfo{publisher}{Springer New York}, \bibinfo{address}{New
  York, NY}, \bibinfo{year}{2014}, pp. \bibinfo{pages}{527--549}. \URLprefix
  \url{https://doi.org/10.1007/978-1-4614-7535-4_22}.
  \DOIprefix\doi{10.1007/978-1-4614-7535-4_22}.
\bibitem[{EGI(2022)}]{checkin}
\bibinfo{author}{EGI}, \bibinfo{title}{{Check-in guide for Service Providers}},
  \bibinfo{howpublished}{\url{https://docs.egi.eu/providers/check-in/sp/}},
  \bibinfo{year}{2022}.
\bibitem[{INSTRUCT-ERIC(2021)}]{InstructERIC}
\bibinfo{author}{INSTRUCT-ERIC}, \bibinfo{title}{Instruct eric structural
  biology web site}, \bibinfo{howpublished}{\url{https://instruct-eric.eu/}},
  \bibinfo{year}{2021}.
\bibitem[{Pablo~Orviz(2022)}]{jepl}
\bibinfo{author}{S.~B. Pablo~Orviz}, \bibinfo{title}{{Jenkins pipeline library
  - official documentation}},
  \bibinfo{howpublished}{\url{https://indigo-dc.github.io/jenkins-pipeline-library/2.0.0/index.html}},
  \bibinfo{year}{2022}.
\bibitem[{Asorey et~al.(2018)Asorey, N{\'u}{\~n}ez, and
  Su{\'a}rez-Dur{\'a}n}]{Asorey2018}
\bibinfo{author}{H.~Asorey}, \bibinfo{author}{L.~A. N{\'u}{\~n}ez},
  \bibinfo{author}{M.~Su{\'a}rez-Dur{\'a}n},
\newblock \bibinfo{title}{{Preliminary Results From the Latin American Giant
  Observatory Space Weather Simulation Chain}},
\newblock \bibinfo{journal}{Space Weather} \bibinfo{volume}{16}
  (\bibinfo{year}{2018}) \bibinfo{pages}{461--475}.
  \DOIprefix\doi{10.1002/2017SW001774}.
\bibitem[{Rubio-Montero et~al.(2021)Rubio-Montero, Pag{\'a}n-Muñoz,
  Mayo-Garc{\'i}a, Pardo-D{\'i}az, Sildelnik, and Asorey}]{LAGO-icrc2021}
\bibinfo{author}{A.~J. Rubio-Montero}, \bibinfo{author}{R.~Pag{\'a}n-Muñoz},
  \bibinfo{author}{R.~Mayo-Garc{\'i}a}, \bibinfo{author}{A.~Pardo-D{\'i}az},
  \bibinfo{author}{I.~Sildelnik}, \bibinfo{author}{H.~Asorey},
\newblock \bibinfo{title}{{The EOSC-Synergy cloud services implementation for
  the Latin American Giant Observatory (LAGO)}},
\newblock in: \bibinfo{booktitle}{37th International Cosmic Ray Conference
  (ICRC2021). PoS(ICRC2021)261}, Proceedings of Science (PoS),
  \bibinfo{publisher}{SISSA}, \bibinfo{address}{Berlin, Germany},
  \bibinfo{year}{12-23 July 2021}. \DOIprefix\doi{10.22323/1.395.0261}.
\bibitem[{Caballer et~al.(2015)Caballer, Blanquer, Molt\'{o}, and
  Alfonso}]{10.1007/s10723-014-9296-5}
\bibinfo{author}{M.~Caballer}, \bibinfo{author}{I.~Blanquer},
  \bibinfo{author}{G.~Molt\'{o}}, \bibinfo{author}{C.~Alfonso},
\newblock \bibinfo{title}{Dynamic management of virtual infrastructures},
\newblock \bibinfo{journal}{J. Grid Comput.} \bibinfo{volume}{13}
  (\bibinfo{year}{2015}) \bibinfo{pages}{53–70}. \URLprefix
  \url{https://doi.org/10.1007/s10723-014-9296-5}.
  \DOIprefix\doi{10.1007/s10723-014-9296-5}.
\bibitem[{Australia(2022)}]{DEA}
\bibinfo{author}{G.~Australia}, \bibinfo{title}{{DEA coastlines}},
  \bibinfo{howpublished}{\url{https://cmi.ga.gov.au/data-products/dea/581/dea-coastlines}},
  \bibinfo{year}{Accessed, January 2022}.
\bibitem[{Bishop-Taylor et~al.(2019)Bishop-Taylor, Sagar, Lymburner, Alam, and
  Sixsmith}]{rs11242984}
\bibinfo{author}{R.~Bishop-Taylor}, \bibinfo{author}{S.~Sagar},
  \bibinfo{author}{L.~Lymburner}, \bibinfo{author}{I.~Alam},
  \bibinfo{author}{J.~Sixsmith},
\newblock \bibinfo{title}{Sub-pixel waterline extraction: Characterising
  accuracy and sensitivity to indices and spectra},
\newblock \bibinfo{journal}{Remote Sensing} \bibinfo{volume}{11}
  (\bibinfo{year}{2019}). \URLprefix
  \url{https://www.mdpi.com/2072-4292/11/24/2984}.
  \DOIprefix\doi{10.3390/rs11242984}.
\bibitem[{Bishop-Taylor et~al.(2021)Bishop-Taylor, Nanson, Sagar, and
  Lymburner}]{BISHOPTAYLOR2021112734}
\bibinfo{author}{R.~Bishop-Taylor}, \bibinfo{author}{R.~Nanson},
  \bibinfo{author}{S.~Sagar}, \bibinfo{author}{L.~Lymburner},
\newblock \bibinfo{title}{Mapping australia's dynamic coastline at mean sea
  level using three decades of landsat imagery},
\newblock \bibinfo{journal}{Remote Sensing of Environment}
  \bibinfo{volume}{267} (\bibinfo{year}{2021}) \bibinfo{pages}{112734}.
  \URLprefix
  \url{https://www.sciencedirect.com/science/article/pii/S0034425721004545}.
  \DOIprefix\doi{https://doi.org/10.1016/j.rse.2021.112734}.
\bibitem[{Copernicus(2022)}]{Dias}
\bibinfo{author}{Copernicus}, \bibinfo{title}{{DIAS services}},
  \bibinfo{howpublished}{\url{https://www.copernicus.eu/en/access-data/dias}},
  \bibinfo{year}{Accessed, January 2022}.
\bibitem[{Mu et~al.(2011)Mu, Zhao, and Running}]{MU20111781}
\bibinfo{author}{Q.~Mu}, \bibinfo{author}{M.~Zhao}, \bibinfo{author}{S.~W.
  Running},
\newblock \bibinfo{title}{Improvements to a modis global terrestrial
  evapotranspiration algorithm},
\newblock \bibinfo{journal}{Remote Sensing of Environment}
  \bibinfo{volume}{115} (\bibinfo{year}{2011}) \bibinfo{pages}{1781--1800}.
  \URLprefix
  \url{https://www.sciencedirect.com/science/article/pii/S0034425711000691}.
  \DOIprefix\doi{https://doi.org/10.1016/j.rse.2011.02.019}.
\bibitem[{Wan et~al.(2015)Wan, Zhang, Xue, Hong, Hong, and
  Gourley}]{Wan20156485}
\bibinfo{author}{Z.~Wan}, \bibinfo{author}{K.~Zhang}, \bibinfo{author}{X.~Xue},
  \bibinfo{author}{Z.~Hong}, \bibinfo{author}{Y.~Hong},
  \bibinfo{author}{J.~Gourley},
\newblock \bibinfo{title}{Water balance-based actual evapotranspiration
  reconstruction from ground and satellite observations over the conterminous
  united states},
\newblock \bibinfo{journal}{Water Resources Research} \bibinfo{volume}{51}
  (\bibinfo{year}{2015}) \bibinfo{pages}{6485--6499}. \URLprefix
  \url{https://www.scopus.com/inward/record.uri?eid=2-s2.0-84941996457&doi=10.1002%2f2015WR017311&partnerID=40&md5=2cb0bc3c5ec45c7de97fb17f45b09ce8}.
  \DOIprefix\doi{10.1002/2015WR017311}, \bibinfo{note}{cited By 56}.
\bibitem[{Goodman et~al.(2019)Goodman, BenYishay, Lv, and
  Runfola}]{GOODMAN2019103}
\bibinfo{author}{S.~Goodman}, \bibinfo{author}{A.~BenYishay},
  \bibinfo{author}{Z.~Lv}, \bibinfo{author}{D.~Runfola},
\newblock \bibinfo{title}{Geoquery: Integrating hpc systems and public
  web-based geospatial data tools},
\newblock \bibinfo{journal}{Computers \& Geosciences} \bibinfo{volume}{122}
  (\bibinfo{year}{2019}) \bibinfo{pages}{103--112}. \URLprefix
  \url{https://www.sciencedirect.com/science/article/pii/S0098300418305326}.
  \DOIprefix\doi{https://doi.org/10.1016/j.cageo.2018.10.009}.
\bibitem[{Abouali et~al.(2013)Abouali, Timmermans, Castillo, and
  Su}]{ABOUALI2013134}
\bibinfo{author}{M.~Abouali}, \bibinfo{author}{J.~Timmermans},
  \bibinfo{author}{J.~E. Castillo}, \bibinfo{author}{B.~Z. Su},
\newblock \bibinfo{title}{A high performance gpu implementation of surface
  energy balance system (sebs) based on cuda-c},
\newblock \bibinfo{journal}{Environmental Modelling \& Software}
  \bibinfo{volume}{41} (\bibinfo{year}{2013}) \bibinfo{pages}{134--138}.
  \URLprefix
  \url{https://www.sciencedirect.com/science/article/pii/S1364815212003106}.
  \DOIprefix\doi{https://doi.org/10.1016/j.envsoft.2012.12.005}.
\bibitem[{Olmedo et~al.(2017)Olmedo, Ortega-Farias, Fonseca-Luengo, de~la
  Fuente-Saiz, and Pe{\~n}ailillo}]{olmedo2017water}
\bibinfo{author}{G.~Olmedo}, \bibinfo{author}{S.~Ortega-Farias},
  \bibinfo{author}{D.~Fonseca-Luengo}, \bibinfo{author}{D.~de~la Fuente-Saiz},
  \bibinfo{author}{F.~Pe{\~n}ailillo},
\newblock \bibinfo{title}{Water: actual evapotranspiration with energy balance
  models},
\newblock \bibinfo{journal}{R Package Version 0.6}  (\bibinfo{year}{2017}).
\bibitem[{Team(2022)}]{googleearth}
\bibinfo{author}{G.~E.~E. Team}, \bibinfo{title}{{Google earth engine: A
  planetary-scale geo-spatial analysis platform.}},
  \bibinfo{howpublished}{\url{https://earthengine.google.com/}},
  \bibinfo{year}{2022}.
\bibitem[{Padarian et~al.(2015)Padarian, Minasny, and
  McBratney}]{PADARIAN201580}
\bibinfo{author}{J.~Padarian}, \bibinfo{author}{B.~Minasny},
  \bibinfo{author}{A.~McBratney},
\newblock \bibinfo{title}{Using google's cloud-based platform for digital soil
  mapping},
\newblock \bibinfo{journal}{Computers \& Geosciences} \bibinfo{volume}{83}
  (\bibinfo{year}{2015}) \bibinfo{pages}{80--88}. \URLprefix
  \url{https://www.sciencedirect.com/science/article/pii/S009830041530008X}.
  \DOIprefix\doi{https://doi.org/10.1016/j.cageo.2015.06.023}.
\bibitem[{Bhatkar(2021)}]{Stion}
\bibinfo{author}{S.~Bhatkar}, \bibinfo{title}{Stion – a software as a service
  for cryo-em data processing on aws},
  \bibinfo{howpublished}{\url{https://aws.amazon.com/blogs/hpc/stion-a-saas-for-cryo-em-data-processing-on-aws}},
  \bibinfo{year}{2021}.
\bibitem[{Cianfrocco et~al.(2018)Cianfrocco, Lahiri, DiMaio, and
  Leschziner}]{Cianfrocco2018}
\bibinfo{author}{M.~A. Cianfrocco}, \bibinfo{author}{I.~Lahiri},
  \bibinfo{author}{F.~DiMaio}, \bibinfo{author}{A.~Leschziner},
\newblock \bibinfo{title}{cryoem-cloud-tools: A software platform to deploy and
  manage cryo-em jobs in the cloud},
\newblock \bibinfo{journal}{J. Struct. Biol.} \bibinfo{volume}{203}
  (\bibinfo{year}{2018}) \bibinfo{pages}{230--235}.
\bibitem[{Cianfrocco et~al.(2017)Cianfrocco, Wong-Barnum, Youn, Wagner, and
  Leschziner}]{Cianfrocco2017}
\bibinfo{author}{M.~A. Cianfrocco}, \bibinfo{author}{M.~Wong-Barnum},
  \bibinfo{author}{C.~Youn}, \bibinfo{author}{R.~Wagner},
  \bibinfo{author}{A.~Leschziner},
\newblock \bibinfo{title}{Cosmic2: A science gateway for cryo-electron
  microscopy structure determination},
\newblock in: \bibinfo{booktitle}{Practice and Experience in Advanced Research
  Computing 2017.}, \bibinfo{year}{2017}, pp. \bibinfo{pages}{1--5}.
\bibitem[{Ferreira et~al.(2020)}]{WorkflowHub}
\bibinfo{author}{R.~Ferreira}, et~al.,
\newblock \bibinfo{title}{Workflowhub: Community framework for enabling
  scientific workflow research and development},
\newblock \bibinfo{journal}{IEEE}  (\bibinfo{year}{2020}). \URLprefix
  \url{http://dx.doi.org/10.1109/WORKS51914.2020.00012}.
  \DOIprefix\doi{10.1109/WORKS51914.2020.00012}.
\bibitem[{DataCite(????)}]{DataCite}
\bibinfo{author}{DataCite}, \bibinfo{title}{Locate, identify, and cite research
  data with the leading global provider of dois for research data.},
  \bibinfo{howpublished}{\url{https://datacite.org/}}, ????
\bibitem[{PalmaEmail et~al.(2014)}]{ROHub}
\bibinfo{author}{R.~PalmaEmail}, et~al., \bibinfo{title}{ROHub — A Digital
  Library of Research Objects Supporting Scientists Towards Reproducible
  Science}, \bibinfo{year}{2014}. \URLprefix
  \url{https://link.springer.com/chapter/10.1007/978-3-319-12024-9_9}.
  \DOIprefix\doi{10.1007/978-3-319-12024-9_9}.
\bibitem[{EUDAT(2021)}]{B2DROP}
\bibinfo{author}{EUDAT}, \bibinfo{title}{B2drop: Sync and share research data},
  \bibinfo{howpublished}{\url{https://eudat.eu/services/userdoc/b2drop}},
  \bibinfo{year}{2021}.
\bibitem[{Asorey et~al.(2016)Asorey, N{\'u}{\~n}ez, Su{\'a}rez-Dur{\'a}n,
  Torres-Niño, Rodr{\'i}guez-Pascual, Rubio-Montero, and
  Mayo-Garc{\'i}a}]{Asorey2016}
\bibinfo{author}{H.~Asorey}, \bibinfo{author}{L.~A. N{\'u}{\~n}ez},
  \bibinfo{author}{M.~Su{\'a}rez-Dur{\'a}n}, \bibinfo{author}{L.~A.
  Torres-Niño}, \bibinfo{author}{M.~Rodr{\'i}guez-Pascual},
  \bibinfo{author}{A.~J. Rubio-Montero}, \bibinfo{author}{R.~Mayo-Garc{\'i}a},
\newblock \bibinfo{title}{{The Latin American Giant Observatory: A Successful
  Collaboration in Latin America Based on Cosmic Rays and Computer Science
  Domains}},
\newblock in: \bibinfo{booktitle}{16th IEEE/ACM International Symposium on
  Cluster, Cloud and Grid Computing (CCGrid)}, \bibinfo{publisher}{IEEE},
  \bibinfo{address}{Cartagena, Colombia}, \bibinfo{year}{16-19 May 2016}, pp.
  \bibinfo{pages}{707--711}. \DOIprefix\doi{10.1109/CCGrid.2016.110}.
\bibitem[{Rodr{\'i}guez-Pascual et~al.(2015)Rodr{\'i}guez-Pascual, LaRocca,
  Kanellopoulo, Carrubba, Inserra, Ricceri, Asorey, Rubio-Montero,
  N{\'u}{\~n}ez-Gonz{\'a}lez, N{\'u}{\~n}ez, Prnjat, Barbera, and
  Mayo-Garc{\'i}a}]{Rodriguez2015}
\bibinfo{author}{M.~Rodr{\'i}guez-Pascual}, \bibinfo{author}{G.~LaRocca},
  \bibinfo{author}{C.~Kanellopoulo}, \bibinfo{author}{C.~Carrubba},
  \bibinfo{author}{G.~Inserra}, \bibinfo{author}{R.~Ricceri},
  \bibinfo{author}{H.~Asorey}, \bibinfo{author}{A.~J. Rubio-Montero},
  \bibinfo{author}{E.~N{\'u}{\~n}ez-Gonz{\'a}lez}, \bibinfo{author}{L.~A.
  N{\'u}{\~n}ez}, \bibinfo{author}{O.~Prnjat}, \bibinfo{author}{R.~Barbera},
  \bibinfo{author}{R.~Mayo-Garc{\'i}a},
\newblock \bibinfo{title}{{A Resilient Methodology for Accessing and Exploiting
  Data and Scientific Codes on Distributed Environments}},
\newblock in: \bibinfo{booktitle}{18th IEEE International Conference on
  Computational Science and Engineering (CSE)}, \bibinfo{publisher}{IEEE},
  \bibinfo{address}{Porto, Portugal.}, \bibinfo{year}{21-23 Oct. 2015}, pp.
  \bibinfo{pages}{319--323}. \DOIprefix\doi{10.1109/CSE.2015.27}.
\bibitem[{Cand et~al.(2020)Cand, Jand, Tanner, and Cheng}]{msprot}
\bibinfo{author}{C.~Cand}, \bibinfo{author}{H.~Jand},
  \bibinfo{author}{J.~Tanner}, \bibinfo{author}{J.~Cheng},
\newblock \bibinfo{title}{Bioinformatics methods for mass spectrometry-based
  proteomics data analysis},
\newblock \bibinfo{journal}{Int J Mol Sci} \bibinfo{volume}{21}
  (\bibinfo{year}{2020}) \bibinfo{pages}{2873}.
  \DOIprefix\doi{10.3390/ijms21082873}.
\bibitem[{Yi et~al.(2016)Yi, Dong, Yun, Deng, Ren, Liu, and Liang}]{YI201617}
\bibinfo{author}{L.~Yi}, \bibinfo{author}{N.~Dong}, \bibinfo{author}{Y.~Yun},
  \bibinfo{author}{B.~Deng}, \bibinfo{author}{D.~Ren},
  \bibinfo{author}{S.~Liu}, \bibinfo{author}{Y.~Liang},
\newblock \bibinfo{title}{Chemometric methods in data processing of mass
  spectrometry-based metabolomics: A review},
\newblock \bibinfo{journal}{Analytica Chimica Acta} \bibinfo{volume}{914}
  (\bibinfo{year}{2016}) \bibinfo{pages}{17--34}.
  \DOIprefix\doi{https://doi.org/10.1016/j.aca.2016.02.001}.
\bibitem[{Guitton et~al.(2017)Guitton, Tremblay-Franco, Corguillé, Martin,
  Pétéra et~al.}]{wf4mb}
\bibinfo{author}{Y.~Guitton}, \bibinfo{author}{M.~Tremblay-Franco},
  \bibinfo{author}{G.~L. Corguillé}, \bibinfo{author}{J.-F. Martin},
  \bibinfo{author}{M.~Pétéra}, et~al.,
\newblock \bibinfo{title}{Create, run, share, publish, and reference your
  {LC–MS, FIA–MS, GC–MS}, and {NMR} data analysis workflows with the
  {Workflow4Metabolomics 3.0 Galaxy} online infrastructure for metabolomics},
\newblock \bibinfo{journal}{The International Journal of Biochemistry \& Cell
  Biology} \bibinfo{volume}{93} (\bibinfo{year}{2017})
  \bibinfo{pages}{89--–101}. \DOIprefix\doi{10.1016/j.biocel.2017.07.002}.
\bibitem[{Wang et~al.(2016)Wang, Carver, Phelan, Sanchez, Garg, Peng, Nguyen
  et~al.}]{gnps}
\bibinfo{author}{M.~Wang}, \bibinfo{author}{J.~J. Carver},
  \bibinfo{author}{V.~V. Phelan}, \bibinfo{author}{L.~M. Sanchez},
  \bibinfo{author}{N.~Garg}, \bibinfo{author}{Y.~Peng}, \bibinfo{author}{D.~D.
  Nguyen}, et~al.,
\newblock \bibinfo{title}{Sharing and community curation of mass spectrometry
  data with global natural products social molecular networking},
\newblock \bibinfo{journal}{Nature biotechnology} \bibinfo{volume}{34}
  (\bibinfo{year}{2016}). \bibinfo{note}{{PMID}: 27504778}.
\bibitem[{Horai et~al.(2010)}]{massbank}
\bibinfo{author}{H.~Horai}, et~al.,
\newblock \bibinfo{title}{{MassBank}: a public repository for sharing mass
  spectral data for life sciences},
\newblock \bibinfo{journal}{J Mass Spectrom} \bibinfo{volume}{45}
  (\bibinfo{year}{2010}) \bibinfo{pages}{703--14}.
  \DOIprefix\doi{10.1002/jms.1777}.
\bibitem[{P{\'e}rez-Padillo et~al.(2021)P{\'e}rez-Padillo, Morillo, Poyato, and
  Montesinos}]{perez_2021}
\bibinfo{author}{J.~P{\'e}rez-Padillo}, \bibinfo{author}{J.~G. Morillo},
  \bibinfo{author}{E.~C. Poyato}, \bibinfo{author}{P.~Montesinos},
\newblock \bibinfo{title}{Open-source application for water supply system
  management: Implementation in a water transmission system in southern spain},
\newblock \bibinfo{journal}{Water} \bibinfo{volume}{13} (\bibinfo{year}{2021})
  \bibinfo{pages}{3652}. \DOIprefix\doi{10.3390/w13243652}.
\bibitem[{Bayer et~al.(2021)Bayer, Ames, and Cleveland}]{bayer_2021}
\bibinfo{author}{T.~Bayer}, \bibinfo{author}{D.~P. Ames},
  \bibinfo{author}{T.~G. Cleveland},
\newblock \bibinfo{title}{Design and development of a web-based epanet model
  catalogue and execution environment},
\newblock \bibinfo{journal}{Annals of GIS} \bibinfo{volume}{27}
  (\bibinfo{year}{2021}) \bibinfo{pages}{247--260}.
  \DOIprefix\doi{10.1080/19475683.2021.1936171}.
\bibitem[{Kruszy{\'n}ski and Dawidowicz(2020)}]{kruszynski_2020}
\bibinfo{author}{W.~Kruszy{\'n}ski}, \bibinfo{author}{J.~Dawidowicz},
\newblock \bibinfo{title}{Computer modeling of water supply and sewerage
  networks as a tool in an integrated water and wastewater management system in
  municipal enterprises},
\newblock \bibinfo{journal}{Journal of Ecological Engineering}
  \bibinfo{volume}{21} (\bibinfo{year}{2020}) \bibinfo{pages}{261--266}.
  \DOIprefix\doi{10.12911/22998993/117533}.
\bibitem[{Organization)(2018)}]{WMO_2018}
\bibinfo{author}{W.~W.~M. Organization)}, \bibinfo{title}{Scientific assessment
  of ozone depletion: 2018}, \bibinfo{year}{2018}.
\bibitem[{Dhomse et~al.(2018)Dhomse, Kinnison, Chipperfield, Salawitch, Cionni,
  Hegglin, Abraham, Akiyoshi, Archibald, Bednarz, Bekki, Braesicke, Butchart,
  Dameris, Deushi, Frith, Hardiman, Hassler, Horowitz, Hu, J\"ockel, Josse,
  Kirner, Kremser, Langematz, Lewis, Marchand, Lin, Mancini, Mar\'ecal, Michou,
  Morgenstern, O'Connor, Oman, Pitari, Plummer, Pyle, Revell, Rozanov,
  Schofield, Stenke, Stone, Sudo, Tilmes, Visioni, Yamashita, and
  Zeng}]{dhomse_2018}
\bibinfo{author}{S.~S. Dhomse}, \bibinfo{author}{D.~Kinnison},
  \bibinfo{author}{M.~P. Chipperfield}, \bibinfo{author}{R.~J. Salawitch},
  \bibinfo{author}{I.~Cionni}, \bibinfo{author}{M.~I. Hegglin},
  \bibinfo{author}{N.~L. Abraham}, \bibinfo{author}{H.~Akiyoshi},
  \bibinfo{author}{A.~T. Archibald}, \bibinfo{author}{E.~M. Bednarz},
  \bibinfo{author}{S.~Bekki}, \bibinfo{author}{P.~Braesicke},
  \bibinfo{author}{N.~Butchart}, \bibinfo{author}{M.~Dameris},
  \bibinfo{author}{M.~Deushi}, \bibinfo{author}{S.~Frith},
  \bibinfo{author}{S.~C. Hardiman}, \bibinfo{author}{B.~Hassler},
  \bibinfo{author}{L.~W. Horowitz}, \bibinfo{author}{R.-M. Hu},
  \bibinfo{author}{P.~J\"ockel}, \bibinfo{author}{B.~Josse},
  \bibinfo{author}{O.~Kirner}, \bibinfo{author}{S.~Kremser},
  \bibinfo{author}{U.~Langematz}, \bibinfo{author}{J.~Lewis},
  \bibinfo{author}{M.~Marchand}, \bibinfo{author}{M.~Lin},
  \bibinfo{author}{E.~Mancini}, \bibinfo{author}{V.~Mar\'ecal},
  \bibinfo{author}{M.~Michou}, \bibinfo{author}{O.~Morgenstern},
  \bibinfo{author}{F.~M. O'Connor}, \bibinfo{author}{L.~Oman},
  \bibinfo{author}{G.~Pitari}, \bibinfo{author}{D.~A. Plummer},
  \bibinfo{author}{J.~A. Pyle}, \bibinfo{author}{L.~E. Revell},
  \bibinfo{author}{E.~Rozanov}, \bibinfo{author}{R.~Schofield},
  \bibinfo{author}{A.~Stenke}, \bibinfo{author}{K.~Stone},
  \bibinfo{author}{K.~Sudo}, \bibinfo{author}{S.~Tilmes},
  \bibinfo{author}{D.~Visioni}, \bibinfo{author}{Y.~Yamashita},
  \bibinfo{author}{G.~Zeng},
\newblock \bibinfo{title}{Estimates of ozone return dates from
  chemistry-climate model initiative simulations},
\newblock \bibinfo{journal}{Atmospheric Chemistry and Physics}
  \bibinfo{volume}{18} (\bibinfo{year}{2018}) \bibinfo{pages}{8409--8438}.
  \URLprefix \url{https://acp.copernicus.org/articles/18/8409/2018/}.
  \DOIprefix\doi{10.5194/acp-18-8409-2018}.
\bibitem[{Keeble et~al.(2021)Keeble, Hassler, Banerjee, Checa-Garcia, Chiodo,
  Davis, Eyring, Griffiths, Morgenstern, Nowack, Zeng, Zhang, Bodeker, Burrows,
  Cameron-Smith, Cugnet, Danek, Deushi, Horowitz, Kubin, Li, Lohmann, Michou,
  Mills, Nabat, Olivi\'e, Park, Seland, Stoll, Wieners, and Wu}]{keeble_2021}
\bibinfo{author}{J.~Keeble}, \bibinfo{author}{B.~Hassler},
  \bibinfo{author}{A.~Banerjee}, \bibinfo{author}{R.~Checa-Garcia},
  \bibinfo{author}{G.~Chiodo}, \bibinfo{author}{S.~Davis},
  \bibinfo{author}{V.~Eyring}, \bibinfo{author}{P.~T. Griffiths},
  \bibinfo{author}{O.~Morgenstern}, \bibinfo{author}{P.~Nowack},
  \bibinfo{author}{G.~Zeng}, \bibinfo{author}{J.~Zhang},
  \bibinfo{author}{G.~Bodeker}, \bibinfo{author}{S.~Burrows},
  \bibinfo{author}{P.~Cameron-Smith}, \bibinfo{author}{D.~Cugnet},
  \bibinfo{author}{C.~Danek}, \bibinfo{author}{M.~Deushi},
  \bibinfo{author}{L.~W. Horowitz}, \bibinfo{author}{A.~Kubin},
  \bibinfo{author}{L.~Li}, \bibinfo{author}{G.~Lohmann},
  \bibinfo{author}{M.~Michou}, \bibinfo{author}{M.~J. Mills},
  \bibinfo{author}{P.~Nabat}, \bibinfo{author}{D.~Olivi\'e},
  \bibinfo{author}{S.~Park}, \bibinfo{author}{{\O}.~Seland},
  \bibinfo{author}{J.~Stoll}, \bibinfo{author}{K.-H. Wieners},
  \bibinfo{author}{T.~Wu},
\newblock \bibinfo{title}{Evaluating stratospheric ozone and water vapour
  changes in cmip6 models from 1850 to 2100},
\newblock \bibinfo{journal}{Atmospheric Chemistry and Physics}
  \bibinfo{volume}{21} (\bibinfo{year}{2021}) \bibinfo{pages}{5015--5061}.
  \URLprefix \url{https://acp.copernicus.org/articles/21/5015/2021/}.
  \DOIprefix\doi{10.5194/acp-21-5015-2021}.
\bibitem[{{European Open Science Cloud Partnership}(2021)}]{eoscdraft}
\bibinfo{author}{{European Open Science Cloud Partnership}},
  \bibinfo{title}{{Draft proposal for the European Open Science Cloud (EOSC)
  Partnership}}, \bibinfo{howpublished}{\url{shorturl.at/tEIMS}},
  \bibinfo{year}{Accessed, December 2021}.
\bibitem[{Commission et~al.(2020)Commission, for Research, and
  Innovation}]{doi/10.2777/870770}
\bibinfo{author}{E.~Commission}, \bibinfo{author}{D.-G. for Research},
  \bibinfo{author}{Innovation}, \bibinfo{title}{Solutions for a sustainable
  EOSC : a FAIR Lady (olim Iron Lady) report from the EOSC Sustainability
  Working Group}, \bibinfo{publisher}{Publications Office},
  \bibinfo{year}{2020}. \DOIprefix\doi{doi/10.2777/870770}.

\end{thebibliography}

\end{document}